\documentclass[a4paper,onecolumn,11pt,accepted=2020-04-12]{quantumarticle}
\pdfoutput=1
\usepackage[numbers,sort&compress]{natbib}
\usepackage[utf8]{inputenc}
\usepackage[english]{babel}
\usepackage[T1]{fontenc}
\usepackage{amsmath}
\usepackage{amssymb}
\usepackage{xcolor}
\usepackage[linkcolor=blue,citecolor=blue]{hyperref}
\usepackage{braket}
\usepackage{graphicx}
\newcommand{\ketbra}[2]{\ensuremath{\ket{#1}\!\bra{#2}}}
\newcommand{\tr}{\mathrm{tr}}

\begin{document}

\title{Quantum-inspired algorithms for multivariate analysis: from interpolation to partial differential equations}

\author{Juan José García Ripoll}
\affiliation{Institute of Fundamental Physics, Calle Serrano 113b, 28006 Madrid, Spain}
\orcid{0000-0001-8993-4624}
\email{jj.garcia.ripoll@csic.es}
\homepage{http://quinfog.hbar.es}
\maketitle

\begin{abstract}
  In this work we study the encoding of smooth, differentiable multivariate function in quantum registers, using quantum computers or tensor-network representations. We show that a large family of distributions can be encoded as low-entanglement states of the quantum register. These states can be efficiently created in a quantum computer, but they are also efficiently stored, manipulated and probed using Matrix-Product States techniques. Inspired by this idea, we present eight quantum-inspired numerical analysis algorithms, that include Fourier sampling, interpolation, differentiation and integration of partial derivative equations. These algorithms combine classical ideas---finite-differences, spectral methods---with the efficient encoding of quantum registers, and well known algorithms, such as the Quantum Fourier Transform. \emph{When these heuristic methods work}, they provide an exponential speed-up over other classical algorithms, such as Monte Carlo integration, finite-difference and fast Fourier transforms (FFT). But even when they don't, some of these algorithms can be translated back to a quantum computer to implement a similar task.
\end{abstract}

\section{Introduction}
\label{sec:introduction}

Quantum computers use the exponential capacity of a Hilbert space to process information. A quantum computer with $m$ qubits can store $2^m$ complex numbers as the components of the quantum register wavefunction, $\ket{\psi} =\sum_{s=0}^{2^m-1}\psi_s\ket{s}.$ A quantum algorithm creates, manipulates and probes these amplitudes to solve a concrete problem. There exist algorithms where the amplitudes $\psi(s)$ encode a smooth function defined over a volume in $\mathbb{R}^N.$ A common example is storing probability distributions in the quantum register~\cite{Zalka1998,grover2002} and developing algorithms to extract expected values~\cite{montanaro2015} or conditional probabilities~\cite{woerner2019}. With this, it becomes possible to perform valuations of complex financial assets~\cite{rebentrost2018,stamatopoulos2019}, VaR estimates~\cite{egger2019}, and other sophisticated interrogations. Using a similar encoding, one may also address radically different problems, such as solving partial differential equations with finite differences~\cite{cao2013,fillion2018,costa2019} or variational methods~\cite{Lubasch2020}.

In this work we discuss how efficient it is to encode discretized functions in a quantum register.  We find that for certain distributions---smooth differentiable functions with bounded derivatives or bounded spectrum---, the accuracy of the discretization increases exponentially with the number of qubits, while the bipartite entanglement grows slowly or even remains bounded with the problem size. This implies such distributions may be constructed with quasi-local operations and polynomial resources on quantum computers. But it also opens an exciting possibility: a family of quantum-inspired matrix-product state (MPS) techniques that can approximate the low entanglement state of a quantum register and provide new classical (and quantum) algorithms for interpolating, differentiating, Fourier transforming or solving differential equations of such distributions. These techniques work efficiently because of an implicit renormalization where different qubits work with different length scales, in a way that lends itself to efficient interpolation and compression. This idea, analogous to multigrid renormalization method developed in Ref.~\cite{Lubasch2018}, has strong parallelisms to earlier developments in the 2D quantum image processing world~\cite{latorre2005}, gives rise to a performance improvement over earlier techniques based on tensor trains~\cite{grasedyck2013,bachmayr2016} or MPS encodings of mode expansions~\cite{iblisdir2007}.

This paper is structured in three parts. The bulk of the work is preceded by a summary (Section~\ref{sec:summary}) of the main results and heuristic algorithms that are developed in this work. The first section addresses the representation of discretized functions in quantum registers. It presents state-of-the-art techniques (Sect.~\ref{sec:gr-construct}) and new discretizations (Sect.~\ref{sec:other-discretizations}) on an equal footing. We argue that these samplings produce to weakly entangled multi-qubit states (Sect.~\ref{sec:bounds}). This prediction is confirmed numerically for common distributions in 1, 2 and 3 dimensions, using exact simulations of up to 28 qubits. The second part of this work builds on this result, introducing the idea of using matrix-product states (MPS) for representing the low-entanglement encoding of such functions in quantum registers. In other words, we propose using an MPS-encoded (classically simulated) \textit{quantum register} to store, compute, manipulate and interrogate smooth multivariate functions. Section~\ref{sec:quantum-inspired-algorithms} articulates this idea, recalling well known algorithms from the literature and how they specialize for our purposes. With these tools at hand, Section~\ref{sec:analysis} develops new quantum-inspired algorithms for the numerical analysis of multivariate functions. These include the mapping of functions to MPS format (Sect.~\ref{sec:exponential}), Fourier analysis (Sect.~\ref{sec:qft}), interpolation methods (Sect.~\ref{sec:interpolation}), and techniques for approximating derivatives of discretized functions (Sect.~\ref{sec:derivatives}), both through finite-difference (Sect.~\ref{sec:finite-differences}) and the Quantum Fourier Transform (QFT, Sect.~\ref{sec:QFT-derivative}). Sect.~\ref{sec:time-evolution} combines these techniques into higher level algorithms for solving partial derivative equations, demonstrating their performance in the Fokker-Planck model. This work is closed with a discussion of the results, including connections to recent advances in tensor-based numerical analysis, and an outlook of applications.

\section{Summary}
\label{sec:summary}

\begin{table}[t!]
  \centering
  \begin{tabular}{lllll}
    Problem & Algorithm & Type & Cost \\
    \hline \hline
    Expected value & Monte Carlo & C & $\mathcal{O}(1/\varepsilon^2)$\ \cite{montanaro2015} \\
    \quad " & Amplitude estimation & Q & $\mathcal{O}(1/\varepsilon)$\ \cite{montanaro2015} \\
    \quad " & MPS & QI & $\mathcal{O}(N\chi^3\log_2(1/\varepsilon))$\ [cf. \S\ref{sec:expected}] \\
    Fourier transform & QFT & Q & $\mathcal{O}(Nm \log(Nm))$\ \cite{hales2000} \\
    \quad " & FFT & C & $\mathcal{O}(Nm2^{Nm})$\ \cite{press1997} \\
    \quad " & MPS QFT & QI & $\mathcal{O}(Nm \times \text{Simp}_{Nm})$\ [cf. \S\ref{sec:qft}] \\
    Interpolation & Linear $(k=1)$ & C & $\mathcal{O}(2^{Nm})$\ \cite{press1997} \\
    \quad " & MPS Linear $(k=1)$ & QI & $\sim\text{Simp}_{Nm}$ [cf. \S\ref{sec:taylor-interpolation}]\\
    \quad " & FFT & C & $\mathcal{O}(N(m+k)2^{N(m+k)})$ [cf. \S\ref{sec:Fourier-interpolation}]\\
    \quad " & MPS QFT & QI & $\sim \text{QFT}_{N(m+k)}$ [cf. \S\ref{sec:Fourier-interpolation}]\\
    PDE Evolution & Finite differences & C & $\mathcal{O}(T_\text{cgs}2^{2Nm})$\ \cite{press1997}\\
    \quad " & MPS differences & QI & $\mathcal{O}(T_\text{cgs}\times\text{Simp}_{Nm})$ [cf. \S\ref{sec:pde-finite-differences}] \\
    \quad " & FFT method & C & $\mathcal{O}((Nm+1)2^{Nm})$\ \cite{weideman1986,press1997} \\
    \quad " & MPS QFT & QI & $\sim 2\times \text{QFT}_{N(m+k)}$ [cf. \S\ref{sec:time-evolution-qft}]\\
    State construct & GR-like & Q & $\mathcal{O}(Nm\chi^2)$ [cf. \S\ref{sec:bounds}]\\
    \quad " & Explicit wavefunction & C & $\mathcal{O}(2^{Nm})$ \\
    \quad " & MPS & QI & $\mathcal{O}(T_\text{steps}\times\text{Simp}_{Nm})$ [e.g. \S\ref{sec:multiplication}]\\
    \hline
    MPS algorithms & Simplification $(\text{Simp}_{Nm})$ & C & $\mathcal{O}(T_\text{sweeps} Nm 4d^3\chi^3)$\ \cite{orus2014}\\
    \quad " & Expected values & C & $\mathcal{O}(Nm\times 2d\chi^3)$\ \cite{orus2014}\\
    \quad " & MPO-MPS product & C & $\mathcal{O}(Nm(d\chi\chi_f)^2)$\ \cite{orus2014}
  \end{tabular}
  \caption{Algorithms and their costs. We compare the costs of different tasks when working with multivariate functions, from the construction of the state, to the simulation of their evolution. We use the following heuristic values: $\varepsilon,$ desired error bound; $N,$ number of variables; $m\sim \mathcal{O}(\log_2(1/\varepsilon)),$ number of qubits per variable for $2^m$ points in discretization; $d=2,$ physical dimension of qubits; $\chi, \chi_f,$ effective MPS and MPO bond dimensions; $T_\text{sweeps}$ number of iterations in the simplification algorithm (cf. Sect.~\ref{sec:mps-approximation}); $T_\text{cgs},$ number of iterations in conjugate gradient method (cf. Sect.~\ref{sec:pde-finite-differences}). Some of the algorithms give rise to their own heuristic metrics: $\mathrm{Simp}_{K},$ cost of simplifying an MPS with $K$ qubits; $\mathrm{QFT}_{K}$, cost of simulating a QFT using MPS with $K$ qubits. We distinguish classical methods (C), from algorithms for a quantum computer (Q), or classical MPS techniques that imitate the quantum register approach (QI). We also reference the papers or sections in this work where such costs are discussed.}
  \label{tab:algorithms}
\end{table}

Quantum computing has introduced the clever idea of encoding probability distributions in quantum registers. In this encoding, a small number of qubits can store an exponentially large number of function samples. To fix ideas and notation, let us take a function $p(x_1,\ldots,x_N)$ of $N$ variables in a bounded interval. Let us use $m$ qubits to uniformly discretize the domain
\begin{align}
  x_s^{i} &= a_i + \frac{b_i-a_i}{2^m}s_i = a_i + \delta^{i}_m \sum_{k=1}^m\frac{s_i^k}{2^k},\;i=1,2\ldots N.\label{eq:coordinates}
\end{align}
The non-negative integer $s_i$ takes all possible values obtained by grouping $m$ bits $s_i^1s_i^2\cdots s_i^m,$ ordered in decreasing significance. The $N$ integers or $Nm$ bits can be associated to different states of a quantum register, enabling two representations of the function $p.$ The first one assumes a non-negative function,
\begin{equation}
  \ket{p} \propto \sum_{s_1,\ldots,s_N} \sqrt{p(s_1,\ldots,s_N)}\ket{s_1}\otimes\cdots\otimes \ket{s_N},\;s_i\in\{0,1,\ldots 2^{m}-1\},\label{eq:representation-1}
\end{equation}
and associates $p(s)$ to the probability of the state $\ket{s}.$ The second encoding does not make this assumption and maps the distribution directly to the wavefunction
\begin{equation}
  \ket{p} \propto \sum_{s_1,\ldots,s_N} p(s_1,\ldots,s_N)\ket{s_1}\otimes\cdots\otimes\ket{s_N}.
  \label{eq:representation-2}
\end{equation}
Both representations can be extended to situations where the functions are not normalized, just by keeping track of global prefactors.

In this work we argue that both representations are exponentially efficient in many ways. First, the quantum register demands only a logarithmically growing number of qubits $Nm$ to store an exponential amount of weights $2^{Nm}$ in a discretized function. Second, we also need an exponentially small number of qubits $m\sim-\log(\varepsilon)$ to reduce the discretization error below a given tolerance $\varepsilon.$ Third and finally, we find that for smooth, differentiable functions with bounded derivatives, these states have a small amount of entanglement. Indeed, for many distributions of interest we obtain an entanglement scaling $\mathcal{O}(N)$ with the dimension of the problem. We conjecture that this behavior is due to an implicit renormalization that happens in the quantum register, where some bits $s_1^1,s_2^1\ldots$ carry information over long wavelengths and large features, and the least significant bits $s_1^m,s_2^m\ldots$ are efficiently approximated with low-entanglement interpolation methods over those qubits (Sect.~\ref{sec:interpolation}).

These findings suggest that many useful functions and problems can be constructed in a quantum register with polynomial resources, due to the bounded entanglement (see Sect.~\ref{sec:bounds}). However, the same results open the field of \textit{quantum-inspired numerical analysis,} which combines the quantum computing encoding of functions with tensor-network representations and algorithms to manipulate them (see Sect.~\ref{sec:quantum-inspired-algorithms}). This approach suggests new algorithms for integrating probability distributions and computing expected values (Sect.~\ref{sec:expected}), for implementing discrete differentiation (Sect.~\ref{sec:derivatives}), Fourier transform (Sect.~\ref{sec:qft}), interpolation (Sect.~\ref{sec:interpolation}) and for the solution of partial differential equations (Sect.~\ref{sec:time-evolution}).

Table~\ref{tab:algorithms} summarizes the algorithms discussed in this work, focusing on five different problems (i) computation of expectation variables from probability distributions, (ii) construction of probability distributions (see Section\ \ref{sec:exponential}), (iii) Fourier analysis (Section\ \ref{sec:qft}) and (iv) interpolation of discretized functions (Section\ \ref{sec:interpolation},  and (v) solution of partial differential equations (Section\ \ref{sec:time-evolution}). The table pairs the quantum-inspired methods based on matrix-product states (type QI), with alternatives that already exist for quantum computers (type Q) or in the field of classical numerical analysis (type C).
For each method, the table provides an estimate of the algorithm's cost, expressed in terms of well known quantities: desired error bounds $\epsilon,$ number of dimensions of the functions involved $N,$ dimension $d=2$ of quantum register elements, logarithm of the discretization size along each dimension $m,$ estimated bond dimension $\chi$ for the MPS representation, number of steps in the MPS approximation algorithms $T_\mathrm{step}, T_\mathrm{sweeps}.$
In the quantum and classical versions, we provide versions of the bounds that are found in the literature, specialized for the parameters we have just mentioned, together with original of state-of-the-art references. In the case of quantum-inspired numerical analysis, we provide standard heuristic performance metrics based on the literature\ \cite{garcia-ripoll2006, orus2014}. These estimates use parameters, such as the bond dimension $\chi$ or the number of iterations in the approximation algorithm $T_\mathrm{sweeps},$ that are very much problem-dependent.  However, as we will see throughout this work, if entanglement remains bounded throughout the simulations, the MPS-simulated quantum register demands a small bond dimension $\chi,$ and the algorithms provide an exponential speedup over other classical techniques---from finite differences to the highly performant FFT techniques.

\section{Storing multivariate functions in quantum registers}
\label{eq:quantum-register}

\subsection{GR discretization}
\label{sec:gr-construct}

One of the earliest works suggesting the encoding of functions in quantum registers is the work by C. Zalka\ \cite{Zalka1998}, later rediscovered by Grover and Rudolph~\cite{grover2002}. This designed a unitary operator $U_p$ that encodes a probability distribution $p(x)$ in an empty quantum register with $m$ qubits
\begin{align}
  \label{eq:pGR}
  \ket{p_\text{GR}^{(m)}} &:= U_p\ket{0,0,\ldots,0} = \sum_{i=0}^{2^m-1} \sqrt{p_\text{GR}^{(m)}(s)}\ket{s},\;\mbox{with}\;
  &p_\text{GR}^{(m)}(s) =\int_{x_s}^{x_{s+1}}p(u)\mathrm{d}u.
\end{align}
The original construct assumes a random variable $x$ in a bounded interval $[a,b]$ subdivided into $2^m$ smaller intervals, labeled by the quantum register states $\ket{s}=\ket{s^1s^2\cdots s^m}.$

A practical application of this encoded state would be the computation of expected values for any observable or function $f(x).$ This requires engineering an observable $\hat{O}^{(m)}_f$ such that
\begin{equation}
  \bar{f} = \int\! f(x) p(x) \mathrm{d}x \simeq \braket{p_\text{GR}^{(m)}|\hat{O}^{(m)}_f|p_\text{GR}^{(m)}} + \varepsilon_\text{int}.
  \label{eq:expected}
\end{equation}
Typically, we approximate $\hat{O}^{(m)}_f =\sum_s f(x_s)\ketbra{s}{s},$ and apply a uniform discretization~\eqref{eq:coordinates}, to have an integration error that decays exponentially with register size,  $\varepsilon_\text{int} \sim \mathcal{O}(\delta_m).$

As found by A. Montanaro~\cite{montanaro2015}, using amplitude estimation with $U_p$ and the operator $\hat{O}_f,$ one may estimate $\bar{f}$ with a precision that scales better than Monte Carlo algorithms. If the cost of implementing $U_p$ is $T_\text{GR},$ and we aim for a sampling precision $\varepsilon_\text{sample},$ the asymptotic time cost of the ideal amplitude estimation algorithm is roughly
\begin{equation}
  T_\text{QCMC}  = \mathcal{O}\left( T_\text{GR}/\varepsilon_\text{sample}\right).
\end{equation}
This represents a favorable scaling when compared with traditional Monte Carlo, where the sampling uncertainty goes as $\mathcal{O}(\varepsilon_\text{sample}^{-2}),$ but only if the cost of encoding the probability state $T_\text{GR}$ remains small or weakly dependent on the integration error $\varepsilon_\text{int}.$

\begin{figure}[t]
  \centering
  \includegraphics[width=0.9\linewidth]{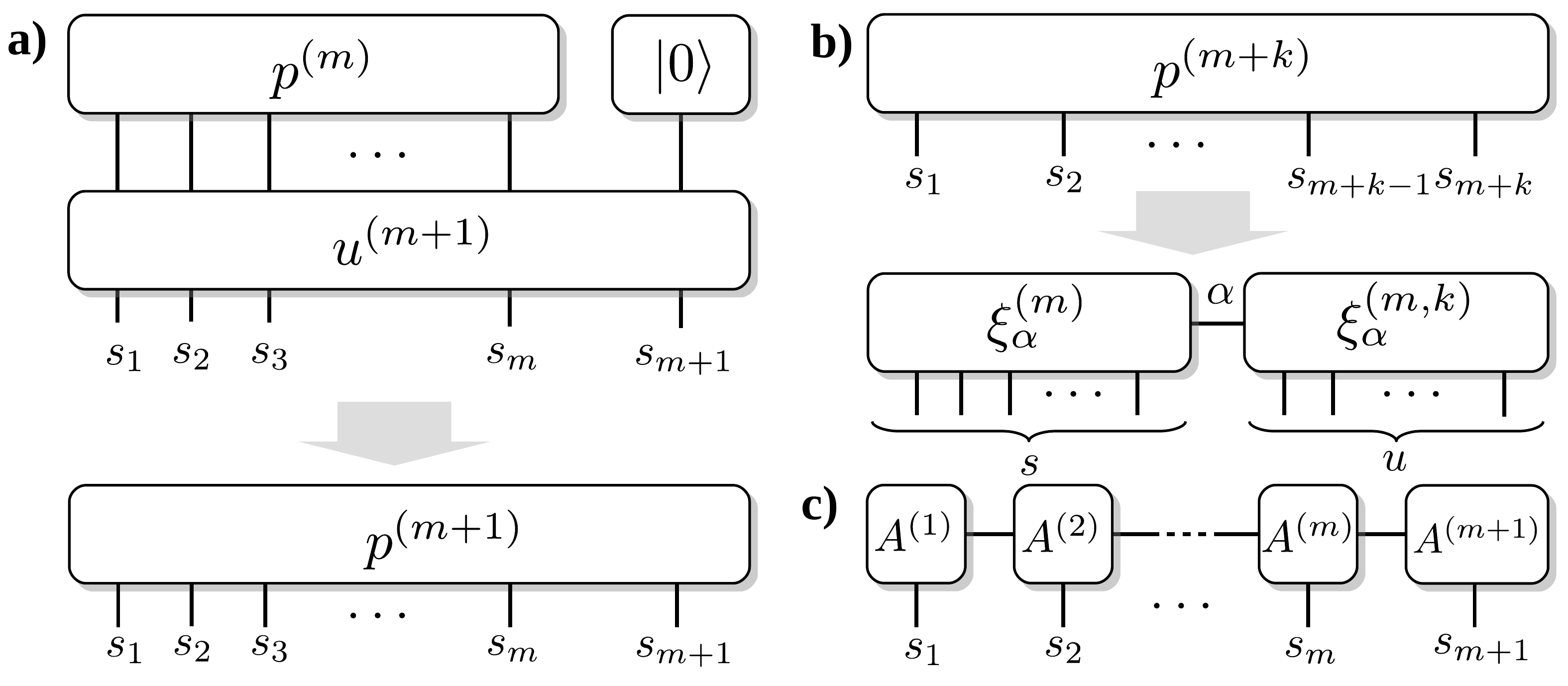}
  \caption{(a) A probability state $\ket{p^{(m)}}$ with discretization size $2^{m+1}$ can be constructed from a coarser state $\ket{p^{(m)}}$ by appending one auxiliary qubit through a unitary operation $U^{(m)}.$ (b) When a state of $m$ qubits is upgraded to $m+k,$ we can study the entanglement between the old and new qubits through a Schmidt decomposition\ \eqref{eq:Schmidt}. (c) If the state has a small Schmidt number for all 1D bipartitions, it is weakly entangled and admits an efficient MPS representation\ \eqref{eq:MPS}.}
  \label{fig:states}
\end{figure}

The $U_p$ is a recursive construct that adds one more qubit of precision in each step.  As sketched in Fig.\ \ref{fig:states}a, the procedure reads
\begin{align}
  \label{eq:GR-unitary}
   \ket{p_\text{GR}^{(m+1)}}&=u^{(m+1)}\ket{p_\text{GR}^{(m)}(s)}\ket{0}
                            = \sum_{s=0}^{2^m-1} \sqrt{p_\text{GR}^{(m)}(s)}\ket{s}(\cos(\theta_s)\ket{0}+\sin(\theta_s)\ket{1})\\
                              &= \sum_{s'=0}^{2^{m+1}-1}\sqrt{p_\text{GR}^{(m+1)}(s')}\ket{s'},
  \nonumber
\end{align}
where we identify $\ket{s}\!\ket{0}=:\ket{2s}$ and $\ket{s}\!\ket{1}=:\ket{2s+1}.$ The rotation angle $0\leq\theta_i\leq\pi/2$ is obtained from two identities
\begin{equation}
  \cos(\theta_s)^2 = \frac{p_\text{GR}^{(m+1)}(2s)}{p_\text{GR}^{(m)}(s)},\; \sin(\theta_s)^2= \frac{p_\text{GR}^{(m+1)}(2s)}{p_\text{GR}^{(m)}(s)}.
\end{equation}
Unfortunately, this algorithm \textit{requires an exponentially large number of angles} and involves a highly non-local unitary with a potentially bad decomposition. The GR algorithm must be therefore considered more a proof of existence, than a practical recipe that can be used when bounding the resources of Monte Carlo analysis.

We will study the GR construct and other discretizations, demonstrating that there are efficient alternatives to equation\ \eqref{eq:GR-unitary}. Our analysis centers on the complexity of the sampled states. Using the bipartite entanglement as quantifier, we will show that the cost of adding one more qubit of resolution decays exponentially. This will help us understand that there one quasi-local unitary procedure that builds the GR state with a cost that is polynomial in the number of qubits, $T_\text{GR}\sim \mathcal{O}(m\log_2(1/\varepsilon)).$ We will confirm numerically this result using various well-known probability distributions.

\subsection{Other discretizations}
\label{sec:other-discretizations}

The integral representation\ \eqref{eq:pGR} encodes the probability that is contained inside each interval, but it requires computing $2^m$ integrals. In practice, this is unnecessary because the estimation of expected values already introduces a discretization error~\eqref{eq:expected} $\varepsilon_\text{sample}\sim \mathcal{O}(\delta_m)$ in the operator definition. It is not difficult to find simpler representations that have the same or better scaling. The obvious one is  the uniform sampling of the probability distribution
\begin{equation}
  \label{eq:riemann}
  \ket{p_\text{R}^{(m)}} = \sqrt{\frac{\delta_m}{N_m}}\sum_s \sqrt{p(x_s)}\ket{s},\;\mbox{with}\; N_m = \sum_s \delta_mp(x_s).
\end{equation}
The standard error bound for this Riemann-type state is $\varepsilon_\text{sample} \leq \max\left|\frac{d}{dx}(fp)\right| \delta_m,$ which depends on the derivatives of both the sampled observables and probability.

The first order scaling is good enough for the simulations that we will show below, because the interval size decreases exponentially with the number of qubits $m.$ We will therefore stick to the GR states or to Eq.~\eqref{eq:riemann}, unless otherwise noted. However, if we need to save some qubits, we can try variations, such as a probability state that implements the trapezoidal or the Simpson rule
\begin{equation}
  \label{eq:Simpson}
\ket{p_\text{S}^{(m)}} = \frac{1}{S_m^{1/2}}\sum_s \sqrt{\theta_sp(x_s)}\ket{s},\;\theta_s=\left\{\begin{array}{ll}1,&s=0,2^{m}-1\\ 4-(s\,\mathrm{mod}\,2),&\mathrm{else}. \end{array}\right.
\end{equation}
The discretization error of the Simpson state decreases faster with the interval size, $\varepsilon_\text{sample}\sim \mathcal{O}(\delta_m^2).$ This scaling means that the qubits required to achieve a given precision $m\sim \log_2(1/\varepsilon_\text{sample})$ can be half those required by the uniform ansatz. These savings may be interesting in resource-starved architectures, such as NISQ computers, and also in the algorithms to be considered later in Sect.~\ref{sec:quantum-inspired-algorithms}.

\subsection{Entanglement bounds}
\label{sec:bounds}

We can show that adding $k$ new qubits to a one-dimensional GR probability state with $m$ qubits demands a vanishingly small amount of entanglement, that decreases exponentially with $2^m.$ This small amount of entanglement suggests that the GR unitary $U^{(m)}_p$ can be replaced with simpler decomposition in terms of quasi-local gates. The reasoning proceeds as follows. Assuming that the probability distribution is smooth, differentiable and has an upper bound on its derivative,
\begin{equation}
  D_p = \max_x |p'(x)|,
  \label{eq:max-derivative}
\end{equation}
we can obtain upper bounds for the entropy of the extra qubit. For one added qubit, $k=1$ our bound reads
\begin{equation}
  S[\rho^{(m,1)}] \leq \sqrt{2(1-P)} \leq 2\sqrt{D_p}|b-a|2^{-m/2}.
  \label{eq:one-qubit-bound}
\end{equation}
If $k$ is larger, we have roughly
\begin{equation}
  S[\rho^{(m,k)}] \leq \mathcal{O}\left( 2\sqrt{D_p}|b-a|2^{-(m-k)/2} \right).
  \label{eq:k-qubit-bound}
\end{equation}
Note that this argument easily extends to the Riemann-type~\eqref{eq:riemann} or Simpson sampling~\eqref{eq:Simpson}, because the differences between them decays exponentially with the number of qubits.

To prove these bounds, assume that we have  produced a GR state with $m$ qubits and we wish to extend it to include $k$ additional qubits. Our  goal is to show that there exists an Schmidt decomposition of the new state into two sets of $m$ and $k$ qubits
\begin{equation}
  \label{eq:Schmidt}
  \ket{p^{(m+k)}} = \sum_\alpha \lambda_\alpha^{1/2}\ket{\xi_\alpha^{(m)}}\otimes\ket{\xi_\alpha^{(m,k)}},
\end{equation}
with a small amount of entanglement between its components
\begin{equation}
  S[\rho^{(m,k)}] = -\sum_\alpha\lambda_\alpha\log(\lambda_\alpha).
\end{equation}

Our first step is to derive an expression for $\rho^{(m,k)}.$ In our extended state, the coordinates $x$ are a function two numbers of $m$ and $k$ bits, which we label $s$ and $u.$ These numbers combine to form a larger previous index $s$ with $m+k$ bits. More precisely
\begin{equation}
  x \in a+ \delta^{(m)}s+\delta^{(m+k)}u=: r_s + \epsilon_u,\;\left\{
    \begin{array}{l}
      s=0,1,\ldots,2^{m}-1, \\
      u=0,1,\ldots,2^{k}-1.
    \end{array}\right.
\end{equation}
For convenience, we will us split $x_z = r_s + \epsilon_u,$ separating a large and small length scale. We can write
\begin{equation}
  \rho^{(m,k)} = \sum_{s,u,v} \sqrt{p^{(m+k)}(r_s+\epsilon_u)p^{(m+k)}(r_s+\epsilon_{v})}\ketbra{u}{v}.
  \label{eq:reduced-dty}
\end{equation}
Starting with Eq.~\eqref{eq:reduced-dty}, we split the density matrix into
\begin{equation}
  \rho^{(m,k)} = \ketbra{e}{e} + \sigma,
\end{equation}
with the unit vector
\begin{equation}
  \ket{e} = \frac{1}{2^{k/2}}\sum_{u=0}^{2^k-1}\ket{u},
\end{equation}
and the perturbation
\begin{equation}
  \braket{u|\sigma|v} = \sum_s\left[ \sqrt{p^{(m+k)}(r_s+\epsilon_u)p^{(m+k)}(r_s+\epsilon_v)}-\frac{1}{2^k}p^{(m)}(r_s) \right].
\end{equation}
Thanks to the mean-value theorem, we know that $p(x)$ achieves its mean value
\begin{equation}
  \bar{p}^{(m)} = \frac{1}{\delta^{(m)}} p^{(m)}(r_s) = p(\bar{r}_s)
\end{equation}
at some point $\bar{r}_s$ in the interval $[r_s,r_s+\delta^{(m)}].$ We can therefore bound
\begin{equation}
  p^{(m+k)}(r_s+\epsilon_u) = \int_{r_s+\epsilon_u}^{r_s+\epsilon_u+\delta^{(m+k)}}p(w)\mathrm{d}w
  \leq \delta^{(m+k)}\times \left[ p(\bar{r}_s) + D_p \delta^{(m)} \right],
\end{equation}
where $D_p=\max_{a\leq x\leq b}p'(x).$ Using $\delta^{(m+k)}=\delta^{(m)}/2^k,$ we bound
\begin{equation}
  |\sigma_{uv}| \leq \sum_s \frac{1}{2^k} D_p (\delta^{(m)})^2 = D_p(b-a)^2 \frac{1}{2^{m+k}}
\end{equation}

We can now use this bound in various ways. We can for instance study the purity of the state, using  $\tr(\sigma^2)\geq 0$ to obtain
\begin{equation}
P[\rho] = \tr{\rho^2}  = 1 + \tr(\sigma^2) - 2\tr\braket{e|\sigma|e} \geq 1 - 2D_p(b-a)^2 \frac{1}{2^{m}},
\end{equation}
a value that vanishes exponentially with the number of qubits. In the particular case in which $k=1,$ the two eigenvalues of the reduced density matrix are
\begin{equation}
\lambda_\pm = \frac{1}{2}\left( 1\pm\sqrt{2P-1} \right).
\end{equation}
Using the bound
\begin{equation}
  H_2(x) = -x\log_2(x)-(1-x)\log_2(1-x)\leq 2\sqrt{x(1-x)},
\end{equation}
we find that the von Neumann entropy is upper bounded by Eq.~\eqref{eq:one-qubit-bound}.

We can also bound the von Neumann entropy of the reduced density matrix for a larger number of extra qubits. Introducing the trace norm of $\sigma$
\begin{equation}
  T[\sigma] = \frac{1}{2}\Vert\sigma\Vert_1 = \sum_{uv}|\sigma_{uv}| \leq D_p(b-a)^2 \frac{1}{2^{m-k+1}}.
\end{equation}
Manipulating the Fannes-Audenaert bound~\cite{audenaert2007}
\begin{equation}
  |S(\rho) - S(\ketbra{e}{e})| \leq T\log_2(2^k-1) +H_2(T) \leq T (k-e^{-k+1}) + 2\sqrt{T},
\end{equation}
we have that the entropy can be made arbitrarily close to zero as we make $T$ smaller and smaller, by increasing $2^{mc-k+1}.$ For sufficiently small $T,$ we can neglect the first term and write~\eqref{eq:one-qubit-bound}.

\subsection{MPS representation}

The fact that adding every new bit requires a small amount of entanglement suggests that the probability state $\ket{p^{(m)}}$ has an efficient matrix-product state (MPS) representation. A matrix product state is a decomposition of a wavefunction as a contraction of matrices that are labeled by the physical indices of a composite quantum system
\begin{equation}
  \ket{p^{(m)}} = \sum_s \sqrt{p^{(m)}(s)}\ket{s} \simeq \sum A_{\alpha_1}^{s_1}A_{\alpha_1,\alpha_2}^{s_2}\cdots
  A^{s_m}_{\alpha_{m}}\ket{s_1,s_2,\ldots,s_m}.
  \label{eq:MPS}
\end{equation}
Each of the tensors is in general different, and has three indices $A_{\alpha_i,\alpha_{i+1}}^{s_i}\in \mathbb{C}^{2\times \chi_i\times \chi_{i+1}}.$ There is one physical index $s$ of dimension 2, and two \textit{bond dimensions} of sizes $\chi_i$ and $\chi_{i+1}.$ Interestingly, MPS states with a small bond dimension $\chi_i$ have also small von Neumann entropy, $S[\rho^{(i,n-i)}] \leq \mathcal{O}(\chi_i).$

Heuristically, we could argue that the bounds for smooth functions indicate the representability of such functions using weakly entangled MP states. In full rigor, the bounds we obtained are bounds on Renyi entropies $S_\alpha$ with $\alpha>=1.$ This  means that, as shown by N. Schuch et al\ \cite{schuch2008}, the MPS simulability of the states is undetermined. This apparent contradiction is observed in the MPS encoding of images\ \cite{latorre2005}. In this case, there are pictures produce a macroscopic number of extremely small eigenvalues in any Schmidt decomposition. These eigenvalues are too small to make the purity or the von Neumann entropy grow, but are too many for an accurate and efficient MPS representation of the state.

However, as we will see below with practical simulations, this does not happen for many smooth functions. Instead, we find very accurate representations of smooth functions in 1, 2 an 3 dimensions using very small MPS. This has important consequences for both quantum and classical algorithms. First, MPS states have efficient sequential constructs~\cite{schoen2005,schoen2007}  with bounded  time-cost $T_\text{GR}\sim \mathcal{O}(m\chi^2),$ potentially smaller than the GR construct. Such constructs are very relevant for the implementation of quantum numerical analysis in quantum computers. Second, MPS states can be manipulated efficiently in classical computers, leading to a new family of quantum-inspired numerical algorithms introduced in later sections.

\subsection{Numerical study of 1D distributions}
\label{sec:numerics-1d}

\begin{figure}[t]
  \centering
  \includegraphics[width=\linewidth]{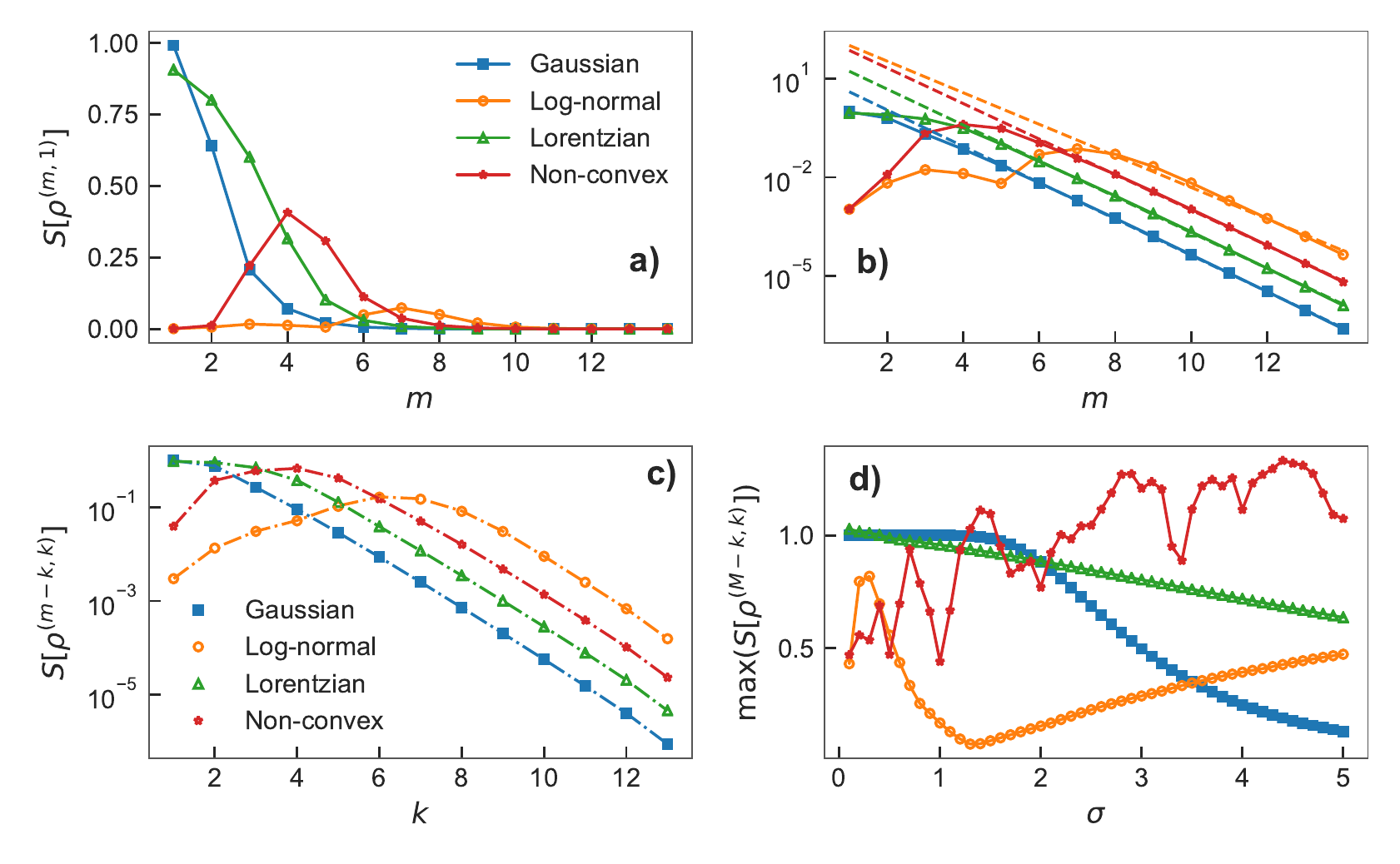}
  \caption{(a) Entropy of entanglement between the $m$ first qubits and an additional qubit that is added to a GR state. We use $(\sigma,\mu)=(1,0),(1,1),(1,0),(1,0)$ for the Gaussian, Log-normal, Lorentzian and non-convex distributions, choosing the intervals $[a,b]=[\mu-6\sigma,\mu+6\sigma],[10^{-16},\mu+50\sigma],\; [\mu-10\sigma,mu+10\sigma]$ and $[0,7\sigma].$ (c-d) Similar as (a), but we plot the entanglement of a bipartition with $(k,M-k)$ qubits, for $M=14.$}
  \label{fig:entropies}
\end{figure}

Let us put these ideas to the test using three paradigmatic distributions
\begin{align}
  \label{eq:Gaussian}
  \mbox{Gaussian:}
  & \quad p_\text{G}(x;\sigma,\mu) = \frac{1}{\sqrt{2\pi}\sigma}e^{-(x-\mu)^2/\sigma^2}, \\
  \label{eq:log-normal}
  \mbox{Log-normal:}
  & \quad p_\text{ln}(x;\sigma,\mu) = \frac{1}{x}p_G(\log(x)),\\
  \label{eq:Lorentzian}
  \mbox{Lorentzian:}
  & \quad p_\text{L}(x;\sigma,\mu) = \frac{\sigma}{2\pi}\frac{1}{(x-\mu)^2+\sigma^2}.
\end{align}
We will also consider an unconventional function that also has a finite bandwidth, but which is not log-convex $\sigma$
\begin{equation}
  \label{eq:non-convex}
  p_\text{nc}(x;\sigma) \propto e^{-x}\sin(2\sigma x)^2\cos(3\sigma x)^2.
\end{equation}
We have chosen all these distributions because we can compute the functions $p^{(m)}(x)$ exactly for all sampling sizes, constructing the GR states for up to $m=14$ qubits\footnote{While $m=14$ qubits do not challenge the computational capabilities of classical computers, in one dimension this precision leads to  discretization error $6\times 10^{-6}$ so small that the plots do not change by enlarging the register.}, analyzing their structure and complexity.

In Fig.\ \ref{fig:entropies}a we plot the entanglement required to enlarge the quantum register by one bit, from $m$ to $m+1.$ This is the entropy $S[\rho^{(m,1)}]$ in the notation above. As shown in Fig.\ \ref{fig:entropies}b, the bipartition never exceeds one \textit{e-bit} of entanglement, and exhibits an exponential decay at large sizes that goes as $2^{-\gamma m}$ with $\gamma$ between 1.73 and 1.84, depending on the simulation and probability distribution. The behavior is therefore more favorable than the bound from Eq.~\eqref{eq:one-qubit-bound}, which overestimates the entanglement.

Fig.\ \ref{fig:entropies}c shows that just like the entropy increase of adding one more qubit is small, the cumulative entropy obtained by studying all bipartitions of $m+k$ qubits also remains small, and with a similar tendency. Moreover, for a detailed enough sampling with $m=14,$ there is no difference between the GR states and the simpler discretizations from Sect.\ \ref{sec:other-discretizations}, shown here with dash-dot lines.

To test whether these favorable dependencies are artifacts of our choice of distributions, we have varied the parameters of the distributions and also tested situations where one of them~\eqref{eq:non-convex} acquires more features. Fig.\ \ref{fig:states}d shows the maximum entanglement entropy over all bipartitions as we change the parameter $\sigma.$ The first three probability distributions have a bounded entanglement below an e-bit. The non-log-convex distribution $p_\text{nc}(x)$ behaves slightly different: increasing $\sigma$ leads to the appearance of more peaks, that are more difficult to describe. This causes a steady increase in entropy, but this is slow enough that still facilitates an efficient construct.

\subsection{Extensions to more variables}
\label{sec:numerics-2d}

Challenging applications to fields such as Physics, fluid dynamics or finance~\cite{orus2019} require the study of states with many more, $p(x_1,x_2\ldots).$ To better understand the scaling of entanglement and the complexity of the state, we have studied the discretization~\eqref{eq:riemann} of two- and three-dimensional Gaussian distributions
\begin{equation}
  \label{eq:Gaussian-nd}
  p(\vec{x}) = \frac{1}{(2\pi)^{N/2}\mathrm{det}(\Sigma)}\exp\left(-\frac{1}{2}\vec{x}^T\Sigma^{-2}\vec{x}\right),\;\vec{x}\in\mathbb{R}^N,
\end{equation}
with covariance matrix $\Sigma$ and zero mean, using the same number of qubits in all dimensions.

As in the 1D problem, we will treat the quantum register as a one-dimensional arrangement of qubits, studying the entanglement over all 1D bipartitions. Naturally, the complexity of this discretization will depend on how we arrange the qubits. The simple straightforward order (A) distributes the qubits sequentially, first by coordinate, then by significance. In this order, Gaussian states with a diagonal covariance matrix $\Sigma=\mathrm{diag}\{\sigma_1,\sigma_2,\ldots\}$ become products states of one-dimensional distributions, such as those studied in Sects.\ \ref{sec:bounds} and \ref{sec:numerics-1d}. We also introduce the order (B), where qubits are first sorted by significance and only then by coordinate. This order is inspired by the renormalization group, and deeply similar to the multi-scale representation of 2D quantum image encodings~\cite{latorre2005}. For a distribution with two random variables, $x_1$ and $x_2,$ using three qubits per variable, the two orders read
\begin{align}
  \mbox{Order (A):}& \ket{s_1s_2} \to \ket{s_1^1}\ket{s_1^2}\ket{s_1^3}\ket{s_2^1}\ket{s_2^2}\ket{s_2^3},
  \;\mbox{and}\label{eq:orders}\\
  \mbox{Order (B):}& \ket{s_1s_2}\to \ket{s_1^1}\ket{s_2^1}\ket{s_1^2}\ket{s_2^2}\ket{s_1^3}\ket{s_2^3}. \nonumber
\end{align}

\begin{figure}[t!]
  \centering
  \includegraphics[width=\linewidth]{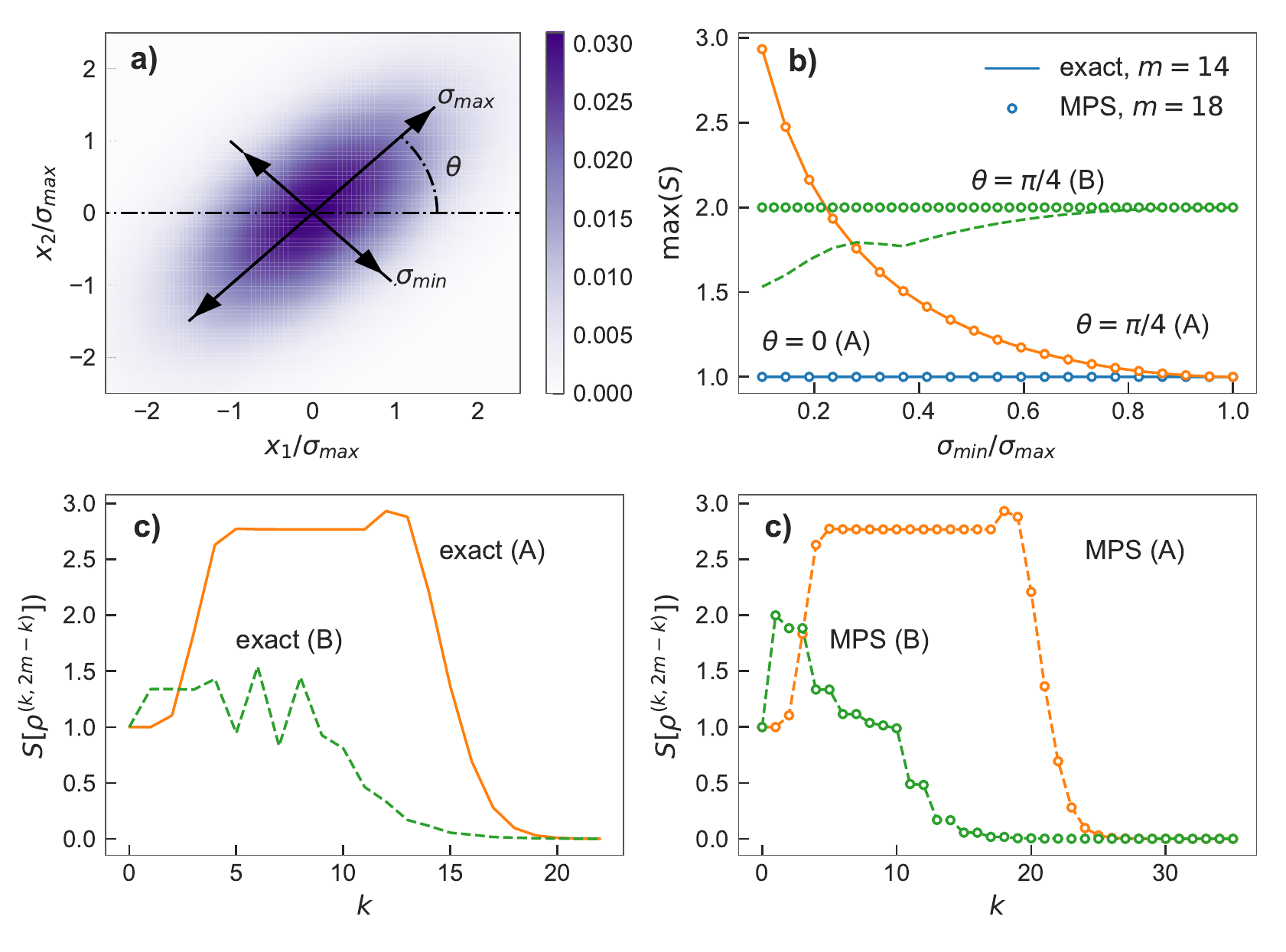}
  \caption{(a) Two-mode Gaussian probability distribution with variances $(\sigma_{\min},\sigma_{\max}),$ rotated an angle $\theta.$ We work with a discretization interval $[-7\sigma_{\max},7\sigma_{\max}]^{\otimes 2},$ compressing $\sigma_{\min}$ down to 0.1, which is a 20dB squeezing of the variance. (b) Maximum entanglement of all bipartitions, for different variances, orientations and orders. In solid lines, we plot the trivial sampling~\eqref{eq:riemann} with $N\times m=28$ qubits in total. We compare those plots with the same simulation using an MPS at a higher sampling (circles). We probe two angles $\theta=0$ and $\pi/4$ in Eq.~\eqref{eq:covariance2d}, and two orders~\eqref{eq:orders}. (c) Entanglement entropies for all bipartitions of $Nn=24$ qubits into $(k,Nm-k),$ for the highly squeezed state $\sigma_{\min}=0.1\sigma_{\max}$ with $\theta=\pi/4.$ (d) Similar plot but for the MPS algorithm using a $Nm=36$ qubits.}
  \label{fig:entropies2d}
\end{figure}

Let us begin the discussion using a general two-dimensional covariance matrix
\begin{equation}
  \Sigma = O(\theta)\left(
    \begin{array}{cc}
      \sigma_{\max} & 0 \\ 0 & \sigma_{\min}
    \end{array}
    \right)O^T,\;
  \mbox{with}\; O =\left(
    \begin{array}{cc}
      \cos(\theta) & \sin(\theta) \\ -\sin(\theta) & \cos(\theta)
    \end{array}
    \right).\label{eq:covariance2d}
\end{equation}
One quadrature is squeezed by a factor $\sigma_{\min}/\sigma_{\max},$ while rotating the frame of reference an angle $\theta$, as shown in Fig.\ \ref{fig:entropies2d}a. We sample this probability, reconstructing the exact wavefunction~\eqref{eq:riemann} with $Nm=28$ qubits. As shown in Table\ \ref{tab:state-sizes}, this size approaches the limits of a decent computer, using 2 gigabytes of data in real double precision form. In contrast, the same distributions using MPS, the techniques from Sect.\ \ref{sec:exponential} and up to $Nm=36$ qubits consume less than 1 megabyte.

\begin{table}[t]
  \centering
  \begin{tabular}{|c|c|c|c|c|c|c|c|}
    \hline
    Discretization & Nm & $\sigma_{\min}/\sigma_{\max}$ & $\theta$ & exact size & order & MPS size \\
    \hline
    $16,384^2$ & 28 & 1 & 0 & $268\times 10^6$ (2 Gb) & (A) & 1,040 (8 kb) \\
    \hline
    " & 28 & 0.1 & $\pi/4$ & " & (A) & 126,628 (1 Mb) \\
    \hline
    " & 28 & 0.1 & $\pi/4$ & " & (B) & 9,388 (71 kb) \\
    \hline
    $262,144^2$ & 36 & 0.1 & $\pi/4$ & $69\times 10^9$ (524 Tb) & (A) & 183,220 (1 Mb) \\
    \hline
    " & 36 & 0.1 & $\pi/4$ & " & (B) & 10,626 (80 kb)\\
    \hline
    \hline
    $2048^3$ & 33 & 0.1 & $\pi/4$ & $8.6\times 10^9$ (64 Gb) & (A) & 825,922 (6.6 Mb) \\
    \hline
   " & 33 & 0.1 & $\pi/4$ & " & (B) & 156,720 (1.22 Mb) \\
    \hline
  \end{tabular}
  \caption{Summary of resources to describe numerically the Gaussian probability distributions with covariance matrices given by~\eqref{eq:covariance2d} and~\eqref{eq:covariance3d}, using $m$ qubits per dimension, either in an exact form---storing all values $p^{(m)}(x_s)$---or in the compact MPS representation~\eqref{eq:MPS}.}
  \label{tab:state-sizes}
\end{table}

The output of these simulations is shown in Fig.~\ref{fig:entropies}. When $\theta=0,$ the wavefunction $\ket{p^{(m)}_\text{2d}}$ is a product state $\ket{p^{(m)}(\sigma_{\max})}\otimes\ket{p^{(m)}(\sigma_{\min})}$ of two one-dimensional distributions. As shown in Fig.\ \ref{fig:entropies}b (blue solid), the maximum entanglement over all one-dimensional bipartitions is less than one e-bit, consistent with Sect.\ \ref{sec:numerics-1d}. To grow the entanglement we must combine squeezing and rotation, recreating a two-mode squeezed state. From the theory of Gaussian states, the entanglement should be maximal for $\theta=\pi/4$ and it should diverge with the squeezing. Our simulations confirm this prediction for the (A) order. Fig.\ \ref{fig:entropies2d}b (orange, solid) shows that the maximum bipartite entanglement between our qubit variables grows as $(\sigma_\text{max}/\sigma_\text{min}){1/4}.$ This entanglement is also spread all along the chain of qubits, as seen in Fig.\ \ref{fig:entropies2d}c.  We can even improve on these results. As shown in Fig.\ \ref{fig:entropies2d}b, if we adopt the renormalization order (B), all states can be described with $N=2$ e-bits of entanglement. Moreover, the entanglement distribution concentrates around the most significant qubits [cf. Fig.\ \ref{fig:entropies2d}c (dashed, green)], producing significantly smaller tensors.

Despite the growth of entanglement, the quantum states that we create always admit a compact MPS representation that beats the classical approach of storing the full wavefunction. As shown in Table\ \ref{tab:state-sizes}, a highly-correlated discretization in the (A) order with 14 and 18 bits per coordinate, requires 1Mb of floating point real numbers in MPS form. The same states stored using the (B) order, take, in the worst scenario $\sigma_{\min}=0.1\sigma_{\max}$ with $\theta=\pi/4,$  just about 80 kilobytes of information. All this is to be compared to the 2 Gb and 524 Tb of data required to write down the wavefunctions of $28$ and $36$ qubits.

\begin{figure}[b!]
  \centering
  \includegraphics[width=0.75\linewidth]{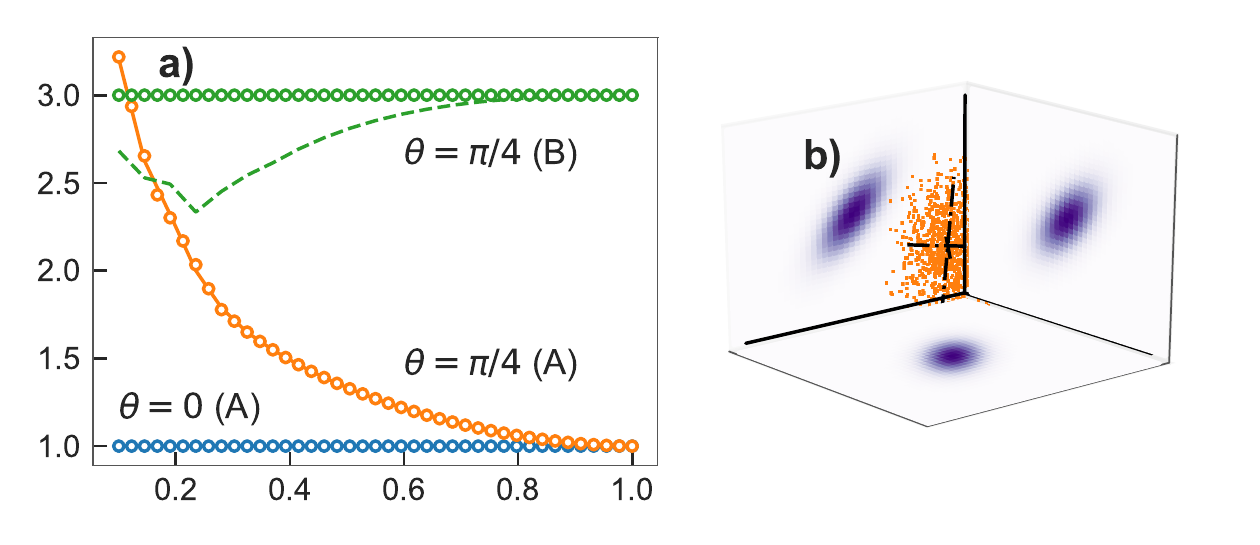}
  \caption{(a) Maximum entanglement over all bipartitions for a three-dimensional Gaussian state~\eqref{eq:covariance3d}, computed with exact wavefunctions ($m=7$ bits per mode, lines) or MPS ($m=9$ bits per mode, circles), using different squeezing rates, angles and orders. (b) Illustration of the three-dimensional Gaussian probability distribution for $\sigma_{\min}=0.5\sigma_{\max}.$}
  \label{fig:entropies3d}
\end{figure}

We have performed the same study using three-dimensional Gaussian states. For concreteness, we have focused on a three-mode squeezed state that starts with a diagonal matrix $\mathrm{diag}(\sigma_{\min},\sigma_{\max},\sigma_{\min})$ and performs two identical rotations around the X and Z axes, with angles $\theta_x=\theta_y=\theta$
\begin{align}
  &\Sigma = O_x(\theta_x)O_z(\theta_y)
    \left(
    \begin{array}{cc}
      \sigma_{\max} & 0 \\
      0 & \sigma_{\min}
    \end{array}
          \right)
          O_z(\theta_z)^TO_x(\theta_x),\;
          \mbox{with}\label{eq:covariance3d}\\
  &O_x =\left(
    \begin{array}{ccc}
      1 & 0 & 0 \\
      0 &  \cos(\theta) & \sin(\theta) \\ 0 & -\sin(\theta) & \cos(\theta)
    \end{array}
    \right),\,O_z =\left(
    \begin{array}{ccc}
      \cos(\theta) & \sin(\theta) & 0 \\ -\sin(\theta) & \cos(\theta) & 0 \\
      0 & 0 & 1
    \end{array}
              \right)
              .\nonumber
\end{align}
We show results for a large, exact simulation with $Nm=21$ qubits, which amounts to $256^3$ points and 128 Mb of data, together with an MPS that is directly built with $33$ qubits, a sampling of $2048^3$ points. As in the two-dimensional case, the combination of squeezing and rotation deforms the state, which as shown in Fig.\ \ref{fig:entropies3d}b becomes a tilted rugby ball.  Once more, the unrotated state remains weakly entangled. It is a product of three one-dimensional probabilities, and the entanglement never exceeds one 1 e-bit [cf. Fig.\ \ref{fig:entropies3d}a (blue, solid)]. The squeezing and rotation leads to a divergence of the maximum entanglement, but this divergence is once more cured by the (B) order. This importance-based structure brings down the entanglement to about 3 e-bits and a reduction of 54,000 in the information required. We conjecture that for these Gaussian states---and other smooth functions---the (B) order consumes at most $N$ e-bits, giving a scaling of resources $\mathcal{O}(2^{2N}\times N)$ in both the time and memory costs of reproducing the probability distribution.

\section{MPS simulation of \textit{quantum registers}}
\label{sec:quantum-inspired-algorithms}

We have seen that it is possible to encode single and multimode probability distributions (and general smooth functions!) in quantum registers, that these states are typically weakly entangled and admit an efficient MPS representation. In the remaining of this work we take this idea a step further, advocating the use of MPS to study multivariate analysis problems, by encoding the solutions to those problems in the state of a fictitious quantum register, and representing and manipulating this register using MPS-simulation techniques.

In order to explain how those algorithms work, we need to review the representation of wavefunctions and operators, and the basic ingredients in the MPS toolbox---tensor contraction and reordering, tensor renormalization, time evolution, etc---that will be used in Sect.~\ref{sec:analysis} to implement actual algorithms. More precisely, Sect.~\ref{sec:bilinear} introduces two representations of multivariate functions: one following the precepts from Sect.~\ref{sec:other-discretizations}, and another one that improves the computation of expected values (Sect.~\ref{sec:expected}) and equation solving. We will also discuss the algorithm of MPS simplification (Sect.~\ref{sec:mps-approximation}), which is an essential tool to implement all numerical analysis approximation schemes. We also refer the reader to a more practical introduction\ \cite{orus2014} and to our SeeMPS library\ \cite{garcia-ripoll2019,garcia-ripoll2020} for details on how these methods are implemented in practice.

\subsection{Quadratic and linear representation}
\label{sec:bilinear}

\begin{figure}[t]
  \centering
  \includegraphics[width=0.8\linewidth]{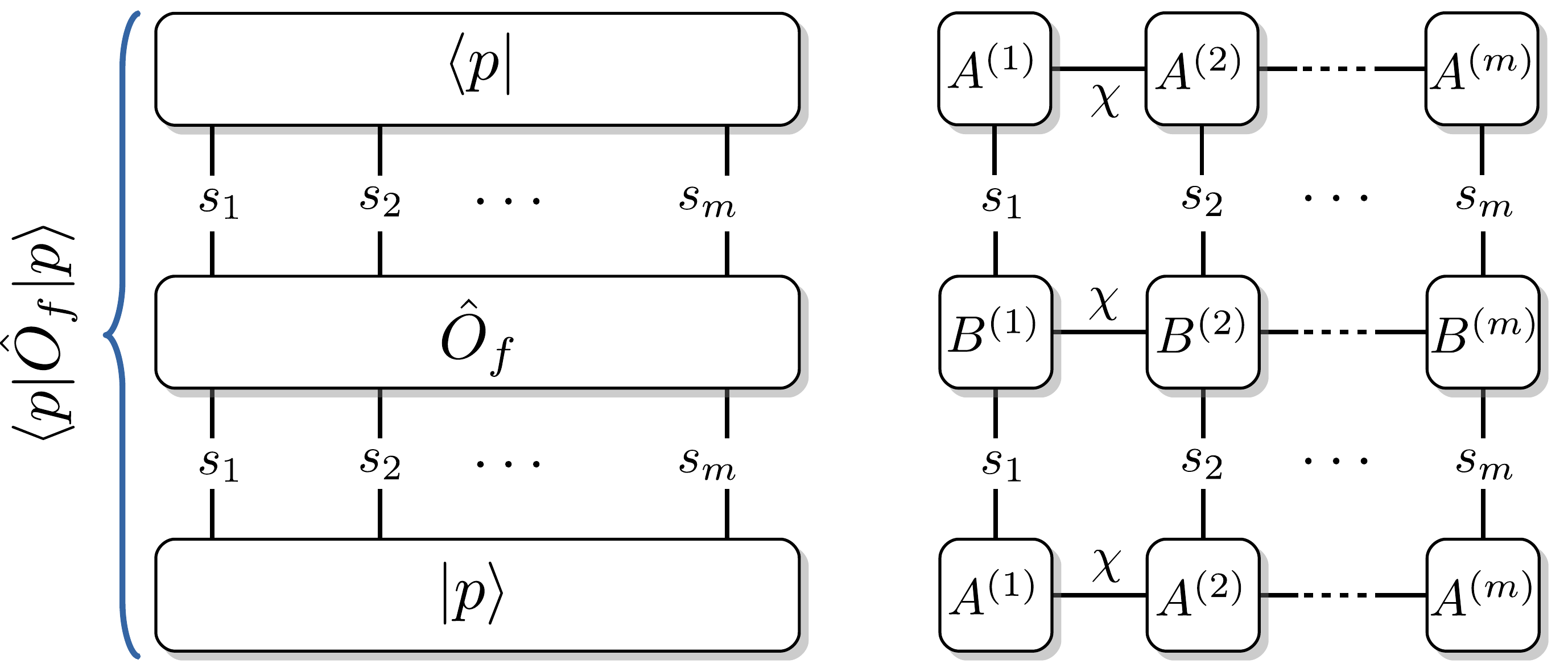}
  \caption{Quadratic MPS representation, where functions are mapped to the amplitude of a wavefunction (as MPS) and observables are mapped to operators (as MPO). Expected values $\bar{f}=\braket{p|\hat{O}_f|p}$ are obtained by contracting three layers of tensors with physical dimensions $d$ and bond dimensions $\sim \chi.$ This has a cost $\mathcal{O}(3d^2\chi^4)$ and causes each tensor to appear twice---in other words, $\bar{f}$ is a quadratic function w.r.t. each tensor.}
  \label{fig:mps-quadratic}
\end{figure}

Our overarching goal for the rest of this work is to encode multivariate functions using a virtual quantum register, and to store this register efficiently as an MPS. If the function we want to encode is non-negative---such as a probability distribution---, we can follow Sect.~\ref{eq:quantum-register}, identify $\sqrt{p}$ with a quantum register wavefunction~\eqref{eq:representation-1} and use the MPS representation~\eqref{eq:MPS} to compress it. In this representation, observables $f(x)$ become Matrix-Product operators (MPO's\ \cite{verstraete2004})
\begin{equation}
  \hat{O}_f^{(m)}= \sum_{\alpha,s} B^{s_1,r_1}_{\beta_1}B^{s_2,r_2}_{\beta_2,\beta_3}\cdots B^{s_m,r_m}_{\beta_m}\ketbra{s_1,s_2\ldots s_m}{r_1,r_2\ldots r_m}.
  \label{eq:mpo}
\end{equation}
We call this the \emph{quadratic MPS representation} because the mean value of an observable $\bar{f}\sim \braket{p|\hat{O}_f|p}$ is a quadratic function of any of the tensors in the MPS state [cf. Fig.~\ref{fig:mps-quadratic}].

In the alternative encoding from equation~\eqref{eq:representation-2}, both observables $f(x)$ and probability distributions $p(x)$ become unnormalized vectors in a Hilbert space
\begin{align}
  \ket{f^{(m)}} &= \sum_{s}f^{(m)}(s)\ket{s} = \sum_{\alpha,s} F^{s_1}_{\beta_1}F^{s_2}_{\beta_2,\beta_3}\cdots F^{s_m}_{\beta_m}\ket{s_1,s_2\ldots s_m},
  \\
  \ket{p^{(m)}}&= \sum_{s}p^{(m)}(s)\bra{s} = \sum_{\alpha,s} A_{\alpha_1}^{s_1}A_{\alpha_1,\alpha_2}^{s_2}\cdots
  A^{s_m}_{\alpha_{m}}\ket{s_1,s_2,\ldots,s_m}.\label{eq:mps-representation-2}
\end{align}
In this representation, each probability state is a linear form that maps observables to expected values, or vice versa
\begin{equation}
  \bar{f} = \sum_s f(s)p(s) = \braket{p|f}\simeq \int\!f(x)p(x)\mathrm{d}x.
  \label{eq:integration-mps}
\end{equation}
We will call this strategy the \emph{linear MPS representation} because $\bar{f}$ is a linear function with respect to any of the tensors in $\ket{p}$ or $\ket{f}.$

\subsection{Memory cost}
\label{sec:memory}

Let us denote $\chi_f,$ $\chi_p$ or simply $\chi$ the largest bond dimensions to encode those functions as MPS. Assuming an $N$-dimensional volume, discretized with $m$ qubits per dimension, the space required by an MPS and an MPO scales as $\mathcal{O}(Nm\chi^2).$ Since $\chi$ dominates this scaling, we need to understand how this dimension behaves in typical problems.

The answer to this question is connected to the bipartite entanglement that is stored in the MPS representations. Since MPS's are obtained through a recursive Schmidt decomposition\ \cite{vidal2003}, the entanglement entropy over any sequential bipartition of the state is bounded by $S\leq \log_2(\chi).$ Conversely, if the maximum entanglement is $S,$ we will expect that the bond dimension scales as $\chi\sim 2^S.$ Thus, for the smooth distributions from Sect.~\ref{sec:numerics-1d}, where $S\sim N,$ we expect a scaling of resources of the form\footnote{This type of cost analysis is standard in the MPS and tensor network literature. You may consult Ref.\ \cite{orus2014} for some general analysis of MPS and MPO contraction and simplification costs.}
\begin{equation}
  \mbox{memory} \sim \mathcal{O}(mN 2^{2S+1}) \sim \mathcal{O}(mN 2^{2N+1})\sim \mathcal{O}\left(N 2^{2N+1}\log_2(1/\varepsilon_\text{int})\right)
\end{equation}
Note that $2^m\sim \varepsilon_\text{int}$ grows with the desired discretization precision. Therefore, if we consider this quantity as ruling parameter, the asymptotic tendency $2^{2N+1}\log_2(1/\varepsilon_\text{int})$ represents an exponential saving over the space $\mathcal{O}(2^{Nm})\sim \mathcal{O}(1/\varepsilon_\text{int}^{N})$ required to store the full wavefunction in general classical algorithms.

This exponential improvement is similar in origin to the one in quantum computers, as it also exploits the rapid growth of the Hilbert space with the number of qubits. We must remark that in the MPS representation the gain is heuristic: it only appears for distributions with nice sampling properties. There are infinitely many problems where this improvement vanishes, due to the growth of $S$ and the exponential blowup of the bond dimensions $\chi.$ However, it also seems that there are still many problems of interest where the MPS representation is useful, as we see below.

\subsection{Integrals and expected values}
\label{sec:expected}

\begin{figure}[t]
  \centering
  \includegraphics[width=0.8\linewidth]{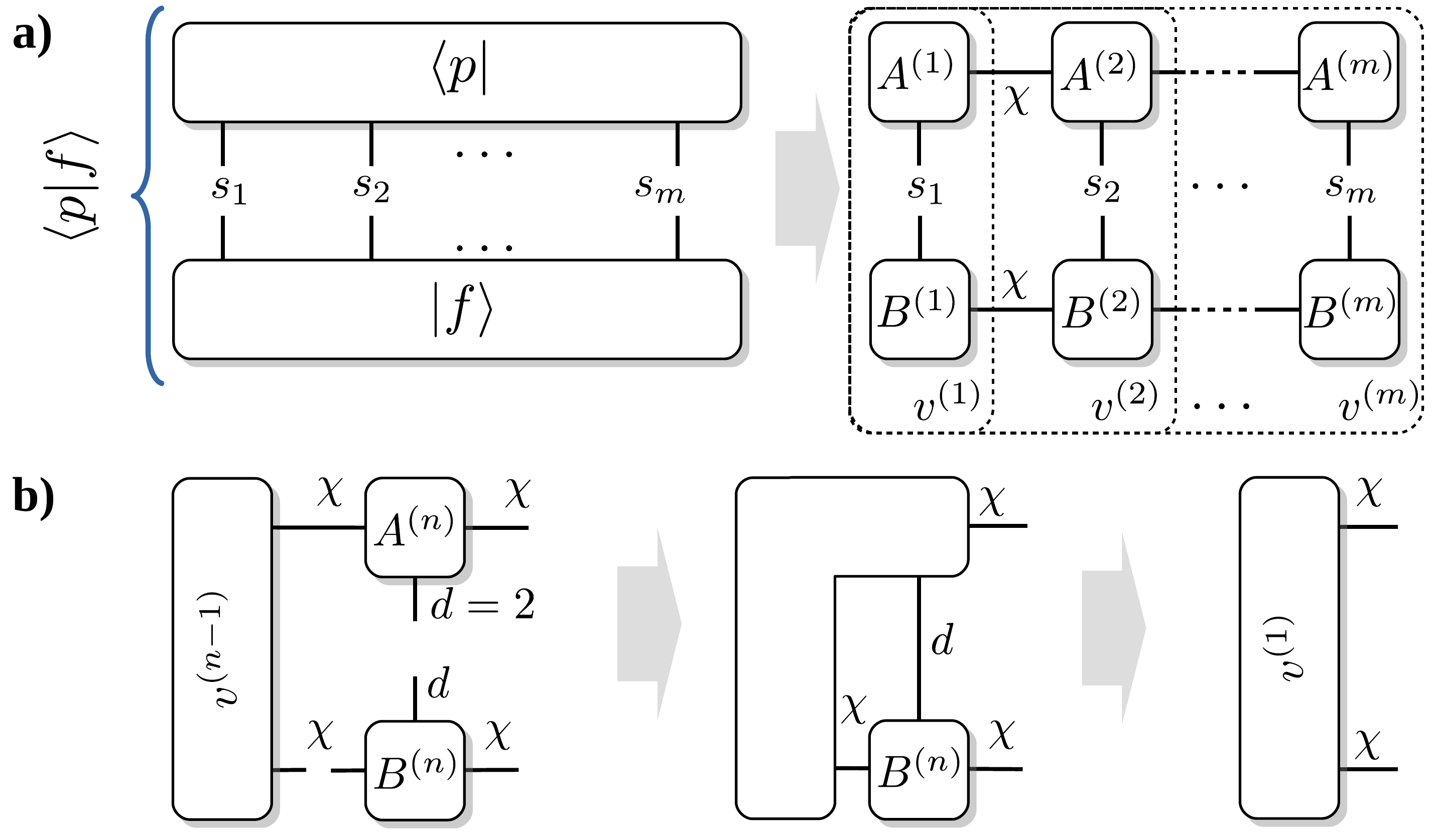}
  \caption{Linear MPS representation. (a) Both observables $\ket{f}$ and a probability distributions $\ket{p}$ have MPS representations with bond dimensions of order $\chi,$ and they combine to give a mean value $\bar{f}=\braket{p|f}$. (b) When there are m qubits, the optimal contraction is sequence of $m\times 2$ steps, with the structure shown here, taking $\mathcal{O}(2d\chi^3)$ operations.}
  \label{fig:mps-linear}
\end{figure}

In the linear MPS representation, the scalar product $\braket{f|p}$ is an approximation to the integral between both functions $\int\!f(x)p(x)\mathrm{d}x.$ As sketched in Fig.~\ref{fig:mps-linear}a, there is an optimal contraction scheme that starts from one boundary---the left-most site in Fig.~\ref{fig:mps-linear}a---, and sequentially contracts with one tensor from $\ket{p}$ and one from $\ket{f}.$ The optimal sequence of contractions in Fig.~\ref{fig:mps-linear}b demands $\mathcal{O}(d\chi^3)$ operations in $2Nm$ steps, with an estimated cost $\mathcal{O}(2dNm\chi^3).$ We can perform a similar analysis for the quadratic MPS representation, where mean values $\bar{f}=\braket{p|\hat{O}_f|p}$ involve contracting the state $\ket{p}$ twice with the MPO $\hat{O}_f$ that represents the observable. In this case the optimal procedure is slightly more involved, with a higher asymptotic cost $\mathcal{O}(Nm\times4\chi^4).$

In order to compare these algorithms with other classical methods, we have to introduce the accuracy of the estimate. In the MPS representation the only error we make is the discretization of $N$ variables with $m$ qubits, which scales as $\varepsilon_\text{int} \sim N2^{m}.$ Thus, the MPS approximation to the integral demands a time $T_\text{constr}+\mathcal{O}(N\log_2(1/\varepsilon_\text{int}/N)\times 2\chi^3),$ where $T_\text{constr}$ is the time to build the MPS and the rest is a logarithmically growing cost associated to the contraction. We can compare this with Monte Carlo sampling\ \cite{press1997}, a good and general method for integration. The errors in this technique arise from the statistical uncertainty $\varepsilon_\text{sample} \sim 1/\sqrt{M}.$ This error decays slowly with the number of iterations $M,$ giving a time cost $\mathcal{O}(1/\varepsilon^2).$ The MPS therefore has the potential of providing an exponential speedup, given that (i) the bond dimension $\chi$ remains small and (ii) the cost of constructing the MPS states is also bounded.

\subsection{Approximating states}
\label{sec:mps-approximation}

In working with the MPS representation, we will frequently need to apply operators that distort and enlarge the tensors. This is corrected by a process known as MPS simplification, which seeks the closest matrix-product state with the smallest bond dimensions, within a prescribed time and error tolerance. The simplification is an optimization typically defined with respect to the 2-norm distance between states
\begin{equation}
  \text{argmin}_{\phi\mbox{ in }\mathrm{MPS}} \Vert \phi - p\Vert^2 = \text{argmin}_\phi d(\phi,p).
\end{equation}
Here $p$ is the state we wish to approximate and $\phi$ is the new MPS. The distance $d(\phi,p)=\braket{p|p} + \braket{\phi|\phi} - 2\mathrm{Re}\braket{\phi|p}$ is a quadratic form with respect to the tensors in $\phi,$ which is optimized iteratively, sweeping across the MPS~\cite{verstraete2008,garcia-ripoll2006,orus2014} in a two-site DMRG-like process.

The cost of this optimization has two parts. The estimation of the linear form $\braket{\phi|p}$ involves a contractions like the ones shown in Fig.~\ref{fig:mps-linear}b, involving $\mathcal{O}(d\chi^3)$ operations per site. On top of this, we apply a two-site simplification algorithm that optimizes pairs of tensors simultaneously, dynamically adapting the bond dimension. This has an extra cost $\mathcal{O}(4(d\chi)^3)$ due to the singular value decomposition. Thus, assuming that we need $T_\text{sweeps}$ for convergence, the simplification time cost grows as $\mathcal{O}(T_\text{sweeps} 4d^3\chi^3).$ In practical examples we find that $T_\text{sweeps}$ is very small: one or two sweeps reach the numerical precision of the computer, giving efficient results.

Finally, note that we can use the algorithm to construct a new state $\phi$ that approximates a linear combination of MPS
\begin{equation}
  \text{argmin}_{\phi\mbox{ in }\mathrm{MPS}} \Vert \phi - \sum_{i=1}^k \alpha_ip_i\Vert^2.
\end{equation}
This has a linear increase in the cost, $\mathcal{O}(kT_\text{sweeps}4d^3\chi^3),$ that has been exploited in other algorithms such as time evolution~\cite{garcia-ripoll2006}.

\subsection{Multiplication by functions and operators}
\label{sec:multiplication}

In many algorithms below we will need to construct a state $\ket{fp}$ that approximates the product of two sampled functions $f(x)p(x).$ This operation can be implemented efficiently in at least four cases, which we now discuss.

Given an arbitrary complex constant $c$ and a discretization of $N$ variables with $m$ qubits per dimension, there exist MPO's for $f(x)=c x,\, c x^2$ and $\exp(c x)$ using bond dimensions $2, Nm$ and $1,$ respectively.  As illustration, let us discuss the implementation of $c x.$ This operator is an MPO~\eqref{eq:mpo} with a bond dimension of size $\chi=2$ that keeps track of whether any operator has been applied. The $n-th$ tensor reads
\begin{equation}
  B^{s_n's_n}_{a,b}=\delta_{s_n',s_n}
  \left\{
    \begin{array}{ll}
      c L 2^{-n}s_n, & a=0,b=1\\
      1, & a=b=1,\;n>1\\
      1, &a=b=0,\\
      0, &\mbox{else}.
    \end{array}
  \right.
\end{equation}

More generally, we can write the exponential of any QUBO formula as a product of MPO's with bond dimension 2
\begin{equation}
  \exp\left(\sum_{i,j=1}^KQ_{ij}s_is_j\right)
  = \prod_{k=1}^K\exp\left( s_i\sum_jQ_{ij}s_j \right).
\end{equation}
The MPO's inside the product are constructed with simple tensors. In particular, for the $k-$th step, the tensors read
\begin{equation}
  B^{s_n's_n}_{a,b}=\delta_{s_n's_n}
  \left\{
    \begin{array}{ll}
      \delta_{a0}\delta_{b0},&n<k,\\
      \delta_{bs_1}\delta_{a0}\exp(Q_{11}s_n), & n=k,\\
      \delta_{ab}\exp(2aQ_{an}s_n), & k<n.
    \end{array}
  \right.\label{eq:QUBO-exponential-tensor}
\end{equation}
Using this idea, we can decompose any exponential or Gaussian, such as $\exp(cx^2+dx)$ or $\exp(\sum_{nm}c_{nm}x_nx_m),$ using $Nm$ layers of MPO's. This technique is discussed in Sect.~\ref{sec:exponential} when we construct MPS.

For other formulas, we find alternative approaches. One is to find an MPO representation of the operator $\hat{O}_f.$ This is expected to be a simple tensor contraction for smooth functions, but the cost of computing it in general may be significant, unless we combine it with strategies such as interpolation (cf. Sect.~\ref{sec:Fourier-interpolation}). Another alternative is to approximate $f(x)$ in Taylor series and use multiplications by constants and by $x,$ combined with simplification stages, to approximate the product. A final approach is to use algorithms from quantum computing to implement the equivalent transformations\ \cite{bhaskar2016,wiebe2016}. The interesting thing in this approach is that, since we are working with artificial registers, we can forego many of the auxiliary qubits, imposing success in all operations.

\section{Multivariate analysis on the MPS-simulated quantum register}
\label{sec:analysis}

After this brief review of MPS method, we will now introduce specific, higher level algorithms for interpolation, Fourier analysis, differentiation, solving PDE's and creating MPS for a sampled function. Unless otherwise stated, all algorithms are implemented using the linear MPS encoding from equations~\eqref{eq:representation-2} and \eqref{eq:mps-representation-2}. Whenever possible we provide also a comparison with the asymptotic performance of state-of-the-art classical algorithms, discussing issues of efficiency, accuracy and numerical stability.

\subsection{Fourier transforms and frequency analysis}
\label{sec:qft}

\begin{figure}[t]
  \centering
  \includegraphics[width=\linewidth]{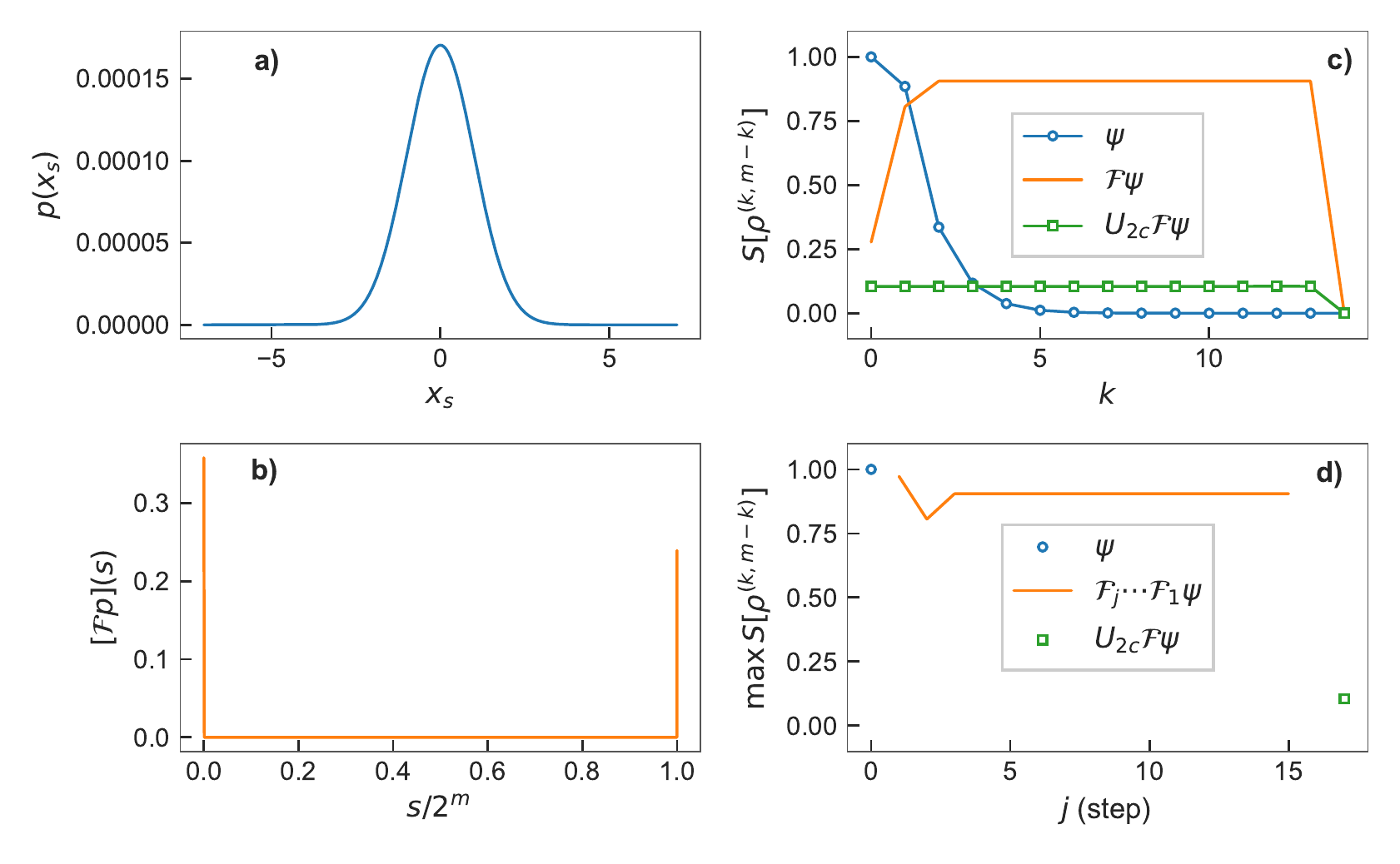}
  \caption{(a) Discretized Gaussian probability distribution with $\sigma=1,\mu=0,$ over 15 qubits, and (b) quantum Fourier transform. (b) Entanglement entropy with respect to all bipartitions of a probability distribution $\psi$ (blue, circles), its quantum Fourier transform $\mathcal{F}\psi$ (orange), and the two's complement of this (green, squares). (c) Maximum entanglement entropy for all bipartitions for the original state (circle), different stages of the QFT (orange) and the QFT after a two's complement transformation.}
  \label{fig:QFT-entropy}
\end{figure}

Assume a function $p(x_s)$ uniformly sampled over $2^m$ points in an interval $x_s\in[a,b]$ of size $L=|b-a|.$ We can define $2^m$ distinct Fourier modes $\tilde\psi_r(x_s)\propto \exp(i q_r x_s)$ for $q_r=2\pi r/L$ and $r=0,1,\ldots 2^m.$ We can also define a Fourier transform
\begin{align}
  \tilde{p}(q_r) = [\mathcal{F}p](q_r)
  &= \frac{1}{2^{m/2}} \sum_{s=0}^{2^m-1}\exp\left( -i q_r (x_s-a) \right)  p(x_s) \label{eq:Fourier}\\
  &= \frac{1}{2^{m/2}} \sum_{s=0}^{2^m-1}\exp\left( -i 2\pi r s/2^m \right)  p(x_s),\nonumber
\end{align}
which returns the expansion of the original function in the Fourier basis. Fortunately, there exists a very efficient classical algorithm, known as the Fast Fourier Transform (FFT), which implements this formula with a cost that is \textit{almost} proportional to the size of our discretrization\ \cite{press1997}, that is $\mathcal{O}(m 2^{m})$ for one dimension or $\mathcal{O}(Nm 2^{Nm})$ for an $N$-dimensional function.

The Quantum Fourier Transform (QFT) is the extension of the classical operator $\mathcal{F}$ to the quantum realm. The QFT  operator $\hat{\mathcal{F}}$ takes a quantum state with wavefunction $p(s)$ and converts it into its Fourier transform
\begin{equation}
  \label{eq:QFT}
  \hat{\mathcal{F}}\ket{p} = \sum_s p(s) \hat{\mathcal{F}}\ket{s} = \sum_r \tilde{p}(r)\ket{r}.
\end{equation}
The QFT is defined by the action on the quantum register states
\begin{equation}
\hat{\mathcal{F}}\ket{s} = \frac{1}{2^{m/2}}\sum_s e^{i2\pi s s'/2^m} \ket{s'}.
\end{equation}
This operation can be implemented using $m+1$ layers of quantum gates, as $\mathcal{F}=\mathcal{U}_\text{flip}\mathcal{F}_{m}\cdots \mathcal{F}_{1}.$ The last operation, $U_\text{flip}$ just reverses the order of the qubits in the quantum register. The $i$-th layer $\mathcal{F}_i$ contains a Hadamard gate $H_i$ on the $i$-th qubit, followed by $m-i$ controlled rotations with respect to the following $m-i$ qubits
\begin{equation}
  \mathcal{F}_i = \prod_{j=i+1}^{m} \exp\left( i\frac{2\pi}{2^{j-i}} \sigma^z_i\otimes \sigma^z_j\right) \exp\left( i\frac{\pi}{2}\sigma^y_i \right).
\end{equation}
Each of the $\mathcal{F}_i$ is an MPO with bond dimension $\chi=2$ and tensors
\begin{align}
  B^{s_n',s_n}_{0,0} &= \delta_{s_n',s_n}, & n< i,\\
  B^{s_i's_i}_{0,s_i'}&=H_{s_i',s_i}, & i\mbox{-th qubit},\notag\\
  B^{s_j',s_j}_{s_i,s_i} & = \exp\left( i\frac{2\pi}{2^{j-i}}s_js_i \right)\delta_{s_j's_j},& j>i.\notag
\end{align}

Applying the QFT on an MPS requires contracting the $Nm$ operators $\mathcal{F}_i$ and simplifying the resulting states. The cost of each simplification stage $\mathrm{Simp}_{Nm}$ dominates the performance of the algorithm, so that we can assume a scaling $\mathcal{O}(Nm\times \mathrm{Simp}_{Nm}).$ The cost of one simplification stage $\mathrm{Simp}_{Nm}\sim \mathcal{O}(T_\text{sweeps}Nm4d^3\chi^3)$ is very much problem dependent, as it depends on the number of required sweeps and entanglement. For smooth functions we expect that the entanglement will be bounded, both because of the estimates above, and because smooth functions will also tend to be concentrated in Fourier space. In that case we can link the cost of one simplification stage $\mathrm{Simp}_{Nm}\sim \mathcal{O}(\log_2(1/\varepsilon_\text{int})$ and of the full QFT $\mathcal{O}(\log_2(1/\varepsilon_\text{int})^2))$  to the precision of the discretization $\varepsilon_\text{int}\propto 2^{-m}.$  This is exponentially faster than the classical FFT $\mathcal{O}(Nm2^{Nm})\sim \mathcal{O}(\log_2(1/\varepsilon_\text{int})/\varepsilon_\text{int}^{N}).$ It is interesting to note that other authors have considered using the QFT in a classical context~\cite{niwa2002,steijl2018}, but they never observed a real speed-up---more like a 20-fold slow down---because of working with the complete wavefunction and not with the MPS-simulated quantum register.

As example, take the Gaussian probability distribution from Fig.~\ref{fig:QFT-entropy}a. Its QFT is a highly concentrated state, another Gaussian in momentum space shown in Fig.~\ref{fig:QFT-entropy}b. We know that the entanglement entropy of the transformed state is upper-bounded by 1 e-bit (see Sect.~\ref{sec:numerics-1d}). This is what we see not only for the final state---orange line in Fig.~\ref{fig:QFT-entropy}c---, but also when we analyze all stages of the transform $\mathcal{F}_1\psi,$ $\mathcal{F}_2\mathcal{F}_1\psi,$ etc [cf. Fig.~\ref{fig:QFT-entropy}d.]

Let us inspect more carefully the Fourier transformed wavefunction Fig.\ \ref{fig:QFT-entropy}b. Note how the wavefunction concentrates on both sides of the interval, $s\simeq 0$ and $s\sim 2^m-1.$ This is caused by the mapping from the non-negative quantum register states $s,$ to the actual momenta, and which has the representation
\begin{equation}
  \hat{k} = \sum_s k_{\bar{s}}\ketbra{s}{s},\;\mbox{with}~
  k_s = \frac{2\pi}{L} \bar{s} \in \left[ \frac{\pi}{L},\frac{\pi}{L} \right].
  \label{eq:momentum}
\end{equation}
Here $\bar{s}$ is the two's complement of the binary number $s=s_1s_2\ldots s_m$
\begin{equation}
  \bar{s} = (1-s_1)\sum_{n=2}^m 2^{m-n}s_n - s_1\left[ 1+ \sum_{n=2}^m 2^{m-n}s_n\right].
\end{equation}
It is useful to implement a two's complement operation that flips all qubits conditioned on the state of the sign qubit, $U_\text{2c}=\ket{s_1,s_2,\ldots,s_m}\to\ket{s_1,s_1\oplus s_2,\ldots,s_1\oplus s_m}.$ When we apply this operator to the Fourier transformed state, $U_\text{2c}\mathcal{F}\psi,$ we find that the amount of entanglement surprisingly drops down, almost close to zero for all bipartitions---see Figs.~\ref{fig:QFT-entropy}c-d. This hints at the fact that the signed quantum register is a much better variable for describing this (and probably other) symmetric probability distributions.

\subsection{Interpolation}
\label{sec:interpolation}

Interpolating means approximating values of a discretized function on points that were not initially considered. We discuss two techniques for interpolating from a quantum register with $m$ qubits to a new register and discretization with $m+k$ qubits. The first method uses finite differences to extrapolate new points as linear combinations of previous values. The second method is a spectral technique based on Fourier transformations that, as we will show, can be exponentially more accurate for finite bandwidth functions.

\subsubsection{Taylor expansions}
\label{sec:taylor-interpolation}

Let us consider a scenario in which we have discretized the interval $[a,b]$ uniformly with $2^m$ points, and we want to add $2^m$ extra points, moving to $m+1$ qubits. Let us call $x_s$ the original variables and $x_{r}'$ the new sampling, which satisfies
\begin{equation}
  x_{2s}'=x_{2s},\; x_{2s+1}'=x_{s}+\delta_m/2.
\end{equation}
We assume that the values at $x_s$ determine those at $x_{2s}'$
\begin{equation}
  p^{(m+1)}(x_{2s}') = p^{(m)}(x_s),\;s=0,1,\ldots 2^{m}-1,\label{eq:linear-interp-1}
\end{equation}
and we only need to extrapolate the values at the odd sites $x_{2s+1}'.$  In order to do so, we can assume that our function is analytic and admits a Taylor expansion to some finite order, which gives the following approximation
\begin{equation}
  p(x_s+\delta_{m+1}) = \frac{1}{2}\left[ p(x_s)+p(x_{s+1}) \right] + \mathcal{O}(\delta_{m+1}^2).\label{eq:linear-interp-2}
\end{equation}
Note that the cost of this classical interpolation algorithm is proportional to the size of the destination vector, thus $\mathcal{O}(2^{N(m+1)})\sim\mathcal{O}(\varepsilon_\text{int}^{-N}),$ considering the precision of the final discretization $\varepsilon_\text{int}\propto 2^{-(m+1)}.$

We can translate equations~\eqref{eq:linear-interp-1} and \eqref{eq:linear-interp-2} into an algorithm that extends a sampling with $m$ qubits into another one with $m+1.$ In a way that resembles very much equation\ \eqref{eq:GR-unitary} , but which is definitely not unitary, we add one least significant qubit at the end
\begin{equation}
  \ket{p^{m+1}} = \ket{p^{m}}\ket{0} + \frac{1}{2}(\hat{S}^-+1)\ket{p^{m}}\ket{1}.
  \label{eq:linear-interpolation}
\end{equation}
The ladder operators $\hat{S}^\pm$ increase or decrease the quantum register by one, displacing the function we encoded
\begin{equation}
  \hat{S}^+ \ket{s} = \left\{ \begin{array}{ll} \ket{s+1},&s<2^m\\ 0, & \mbox{else} \end{array} \right.\quad \hat{S}^-=(\hat{S}^+)^\dagger.
  \label{eq:ladder}
\end{equation}
Instead of using reversible operations (Toffoli, CNOT), we implement the MPO as a smaller, irreversible and classical circuit with one carry bit that propagates through the bond-dimension. This requires a single tensor $C$
\begin{equation}
  \hat{S}^+ = \sum C^{s_1',s_1}_{0,a_1} C^{s_2',s_2}_{a_1,a_2}\cdots C^{s_m',s_m}_{a_m,1}\ket{s_1',s_2'\ldots s_m'}\!\bra{s_1,s_2\ldots s_m}.
\end{equation}
The tensor $C^{s',s}_{a,b}$ is nonzero only for $s'=s\oplus b,$ and $a=s \land b.$ The MPO $\hat{S}^-$ is obtained by simply exchanging the indices as
\begin{equation}
  \hat{S}^- = \sum C^{s_1',s_1}_{0,a_1} C^{s_2',s_2}_{a_1,a_2}\cdots C^{s_m',s_m}_{a_m,1}\ketbra{s_1,s_2\ldots s_m}{s_1',s_2'\ldots s_m'},
\end{equation}
and the linear combination algorithm~\eqref{eq:linear-interpolation} can be implemented using the simplification techniques from Sect.~\ref{sec:mps-approximation}

Heuristically, the MPS algorithm can be exponentially more efficient than trivial interpolation. Since the MPO's have very small bond dimensions, they can be integrated in the simplification phase\ \cite{orus2014}, which dominates the overall cost [cf. Table\ \ref{tab:algorithms}]. More precisely, in the case of smooth functions with low-entanglement representations, the MPS tensors have a small bond dimension and the simplification will require $\mathrm{Simp}_{N(m+1)} \sim \mathcal{O}(\log_2(1/\varepsilon_\text{int}))$ operations, which behaves heuristically better than the scaling of the classical algorithm.

\subsubsection{Fourier interpolation}
\label{sec:Fourier-interpolation}

\begin{figure}[t!]
  \centering
  \includegraphics[width=0.8\linewidth]{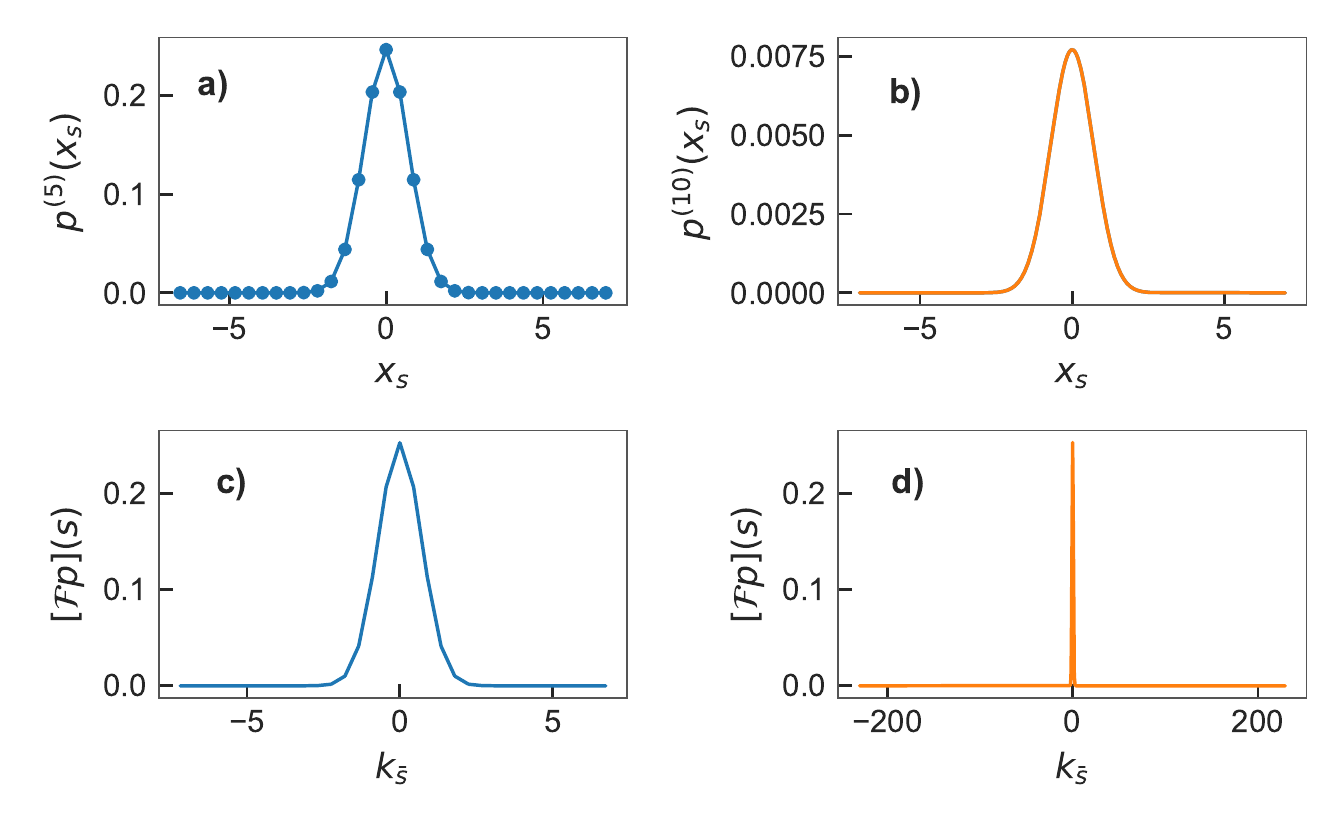}
  \caption{(a) Gaussian probability distribution sampled with $m=5$ qubits and $32$ points. (b) Fourier interpolation and exact sampling with $m=10$ qubits and $1024$ points. Both samplings are indistinguishable. (c) Fourier spectrum of the original distribution. Note how higher frequency components are negligible. (d) Fourier transform of the interpolated function. We have added zero values over $2^5$ sites at higher momenta, outside the axes.}
  \label{fig:QFT-interpolation}
\end{figure}

The accuracy of linear interpolation is constrained by the initial sampling, $\mathcal{O}(\delta_m^2).$ This seems to contradict the Nyquist-Shannon theorem, according to which a function with a bounded spectrum only needs to be sampled with a frequency of $2\times\nu_\text{max}$ for a perfect interpolation. Take for instance a Gaussian probability distribution~\eqref{eq:Gaussian} with width $\sigma,$ which has been sampled in the interval $[-8\sigma,8\sigma].$  Its Fourier transform is a normal distribution with center $\nu=0$ and width $1/\sigma.$ We can say that the information beyond $\nu=4/\sigma,$ is exponentially suppressed. According to Nyquist's theorem, we can exactly reconstruct a Gaussian function by sampling it with period $\sigma/8.$ For the conditions above, that means $8/\sigma\times 16\sigma\sim 128$ points stored in $7$ qubits. However, if we attempt linear interpolation, we typically will make a bounded error that is fixed by the initial sampling, $\mathcal{O}(\delta{x}\sim 1/16).$

A well known solution is to do the interpolation in frequency space. When we perform a discrete Fourier transform, we are decomposing the sampled function $p(x_s)$ as a sum of discrete Fourier modes $\psi(k_{\bar{s}'})\propto \exp(ik_{\bar{s}'}x_s).$ We can use this to reconstruct a \textit{continuous approximation} to the original function $p(x)$ in what is known as \emph{Fourier interpolation}\ \cite{monro1979}
\begin{equation}
  p(x) \simeq \frac{1}{\sqrt{|b-a|}}\sum_{s=0}^{2^{m}-1} \exp(ik_{\bar{s}}x)\left[ \mathcal{F}p^{(m)} \right](k_{\bar{s}}),\;x\in[a,b].
  \label{eq:Fourier-interpolation}
\end{equation}
This continuous approximation can then be resampled with as fine a grid as needed. The technique is so ubiquitous that it is available in standard numerical toolboxes, such as Matlab.

We can implement the approximation and re-sampling very efficiently. If we wish to enlarge the number of qubits from $m$ to $m+k,$ we compute
\begin{equation}
  \ket{p^{(m+k)}}=\hat{\mathcal{F}}^{(m+k)}U_\text{2c}^{m+k}\left[ \left( U_\text{2c}\hat{\mathcal{F}}^{(m)}\ket{p^{(m)}} \right)\otimes \ket{0_2,0_3,\ldots,0_{k+1}} \right].
\end{equation}
We start with a Fourier transform and two's complement over $m$ qubits. We then insert $k$ qubits in positions $2$ to $k+1.$ These are bits that encode very high frequencies and which are populated with zeros, as we do not need any finer details in the sampled function. We finally take the enlarged register and invert both the two's complement and the Fourier transform. As illustrated in Fig.~\ref{fig:QFT-interpolation} this is a powerful technique that can reconstruct a Gaussian using $2^{10}=1024$ points out of a discretized Gaussian with $2^{5}=128$ points, with negligible error.

Note that the classical and MPS-simulated version of Fourier interpolation contain the same ingredients: zero-padding in between two discrete Fourier transforms, which dominate the scaling of the algorithms. However, when we operate with MPS-encoded functions, the improved performance of the QFT with MPS can provide an exponential speedup in practical situations. Following Section\ \ref{sec:qft}, if we extend the discretization, interpolating from a grid with $2^m$ points to one with $2^{m+k},$ the classical algorithm based on the FFT will scale as $\mathcal{O}(N(m+k)2^{N(m+k)})\sim \mathcal{O}(\log_2(1/\varepsilon_\text{int})/\varepsilon_\text{int}^{N})$ if we consider the discretization error of the final lattice $\varepsilon_\text{int}.$ The MPS on the other hand will scale as $\mathcal{O}(N(m+k)\times \mathrm{Simp}_{N(m+k)}).$ If the practical bond dimension remains bounded $\chi,$ this implies a behavior $\mathcal{O}(T_\text{sweeps}N^2(m+k)^2 4d^3\chi^3)\sim \mathcal{O}(\log_2(1/\varepsilon_\text{int})^2 N^2\chi^3).$ This is exponentially faster in many useful situations.

\subsection{Differentiation}
\label{sec:derivatives}

In numerical analysis, there are two main ways to estimate the derivative of a discretized function. The first method is called finite differences, because it relies on linear combinations of the function $p(x_s)$ and its displacements $p(x_s\pm n \delta{x}).$ The second type of methods are called spectral methods, because they work with the Fourier expansion from\ \eqref{eq:Fourier-interpolation}. Both methods have simple translations to the language of MPO's and MPS's.

\subsubsection{Finite differences}
\label{sec:finite-differences}

We illustrate this technique using standard finite-difference approximations to first and second spatial derivatives\ \cite{press1997},
\begin{align}
  &\frac{\partial}{\partial x} p(x)
    \simeq \frac{p(x+\delta{x})-p(x-\delta{x})}{2\delta{x}}+\mathcal{O}(\delta{x}^2), \\
  &\frac{\partial^2}{\partial x^2} p(x)
  \simeq \frac{p(x+\delta{x})p(x-\delta{x})-2p(x)}{\delta{x}^2}+\mathcal{O}(\delta{x}).
\end{align}
Note how the error of the approximation depends on the discretization step $\delta{x}\sim (b-a)/2^m$ of the uniform sampling~\eqref{eq:coordinates}. In classical algorithms, these formulas are applied onto a vector of values representing the function discretized over an $N$-dimensional grid with $2^m$ points along each dimension. In this scenario, the cost of this operation is proportional to the size of the vector that encodes the function $\mathcal{O}(2^{Nm})\sim \mathcal{O}(\varepsilon_\text{int}^{-N}).$

We can improve on this by using the MPS representation, which requires less memory to store the same function. Here, small changes or displacements on the grid $x_s\pm\delta{x}$ map to increments and decrements of the quantum register variable $s.$ These motions are implemented using the ladder operators~\eqref{eq:ladder}. Our finite-difference formulas become
\begin{equation}
  \partial_x\ket{p}  \simeq\frac{\hat{S}^+ - \hat{S}^-}{2\delta{x}}\ket{p},\;
  \partial_{xx}^2\ket{p} \simeq\frac{\hat{S}^+ + \hat{S}^- - 1}{\delta{x}^2}\ket{p}.
  \label{eq:finite-differences-example}
\end{equation}
These formulas can be implemented as MPO's of bond dimension 3, which can be efficiently contracted and simplified using the algorithm from Sect.~\ref{sec:mps-approximation}. Unsurprisingly, as in the linear interpolation algorithm, the cost is dominated by the simplification phase $\mathrm{Simp}_{Nm}\sim \mathcal{O}(\log_2(1/\varepsilon_\text{int})N\chi^3)$, which can be exponentially faster for functions that allow a low-entanglement representation with small bond dimension $\chi.$

Note that higher order approximations are also possible using the same operator or powers of it. Roughly, the bond dimension of the MPO that implements a finite difference formula grows linearly with the order of the approximation, just like the number of non-diagonals in its matrix representation. This is a moderate cost that makes differentiation an approachable routine in higher level algorithms.

\subsubsection{Fourier approximations to derivatives}
\label{sec:QFT-derivative}

Since Fourier interpolation works so well, we can use it to approximate the action derivatives at arbitrarily high orders. Take any differential operator $G(\partial_x)$, such as $\partial_x,\,\partial_x^2,a\partial_x+b\partial_x^2,$ etc. The action of this operator on the Fourier interpolated continuous function~\eqref{eq:Fourier-interpolation} is very simple
\begin{equation}
  G(\partial_x)p(x) \simeq \frac{1}{\sqrt{|b-a|}}\sum_{s=0}^{2^{m}-1} G(ik_{\bar{s}})\exp(ik_{\bar{s}}x)\left[ \mathcal{F}p^{(m)} \right](k_{\bar{s}}),\;x\in[a,b] .
\end{equation}
This formula tells us to Fourier transform the discretized function, multiply it by some factors $G(ik_s)$ and then Fourier-transform back to position space. In a classical scenario, where we store a function using $2^{Nm}$ values, the cost of this algorithm is dominated by the FFT $\mathcal{O}(Nm 2^{Nm})\sim\mathcal{O}(N\log_2(1/\varepsilon_\text{int})/\varepsilon_\text{int}^N).$

We can translate this recipe to our representation of functions in a quantum register. This leads to a non-unitary transformation, with two QFT's and multiplication by a function of the momentum operator~\eqref{eq:momentum}
\begin{equation}
  G(\partial_x)\ket{p^{(m)}} = \hat{\mathcal{F}}^{-1} G(i\hat{k}) \hat{\mathcal{F}}\ket{p^{(m)}}.
\end{equation}
Even though this is a non-unitary operation, it can be computed using the MPS representation. In this case the algorithmic cost is equivalent to that of a QFT, plus the product with the operator $G(i\hat{k}).$ Fortunately, for first and second order derivatives $G(i\hat{k})\sim a \hat{k} + b\hat{k}^2$ has an efficient MPO representation and we revert back to the good heuristic scaling of $\mathrm{Simp}_{Nm}\sim \mathcal{O}(N\chi^3\log_2(1/\varepsilon_\text{int}))$ for smooth functions.

\subsection{Solving partial differential equations}
\label{sec:time-evolution}

One of the main applications of all these techniques is the study of how multivariate functions evolve in time, when subject to one of many partial differential equations. We focus our discussion on the Fokker-Planck equation in one dimension
\begin{equation}
  \frac{\partial}{\partial t}p(x,t) = -\frac{\partial}{\partial x}[\mu(x,t)p(x,t)] + \frac{\partial^2}{\partial x^2}[D(x,t)p(x,t)],
  \label{eq:FP}
\end{equation}
using uniform drift and diffusion, $\mu(x,t)=\mu$ and $D(x,t)=D.$ This equation governs the evolution of probability distributions for random variables undergoing a Wiener process in the Îto representation---a recurrent problem in quantum optics and finance, for instance. Moreover, this is already a challenging toy model from numerical analysis that is subject to numerical instabilities and demands state-of-the-art integration techniques. We will provide two techniques to solve this equation using the MPS encoding of a quantum register, matching the two methods to work with differential operators from Sects.~\ref{sec:finite-differences} and~\ref{sec:QFT-derivative}.

\subsubsection{Finite differences}
\label{sec:pde-finite-differences}

\begin{figure}[t!]
  \centering
  \includegraphics[width=\linewidth]{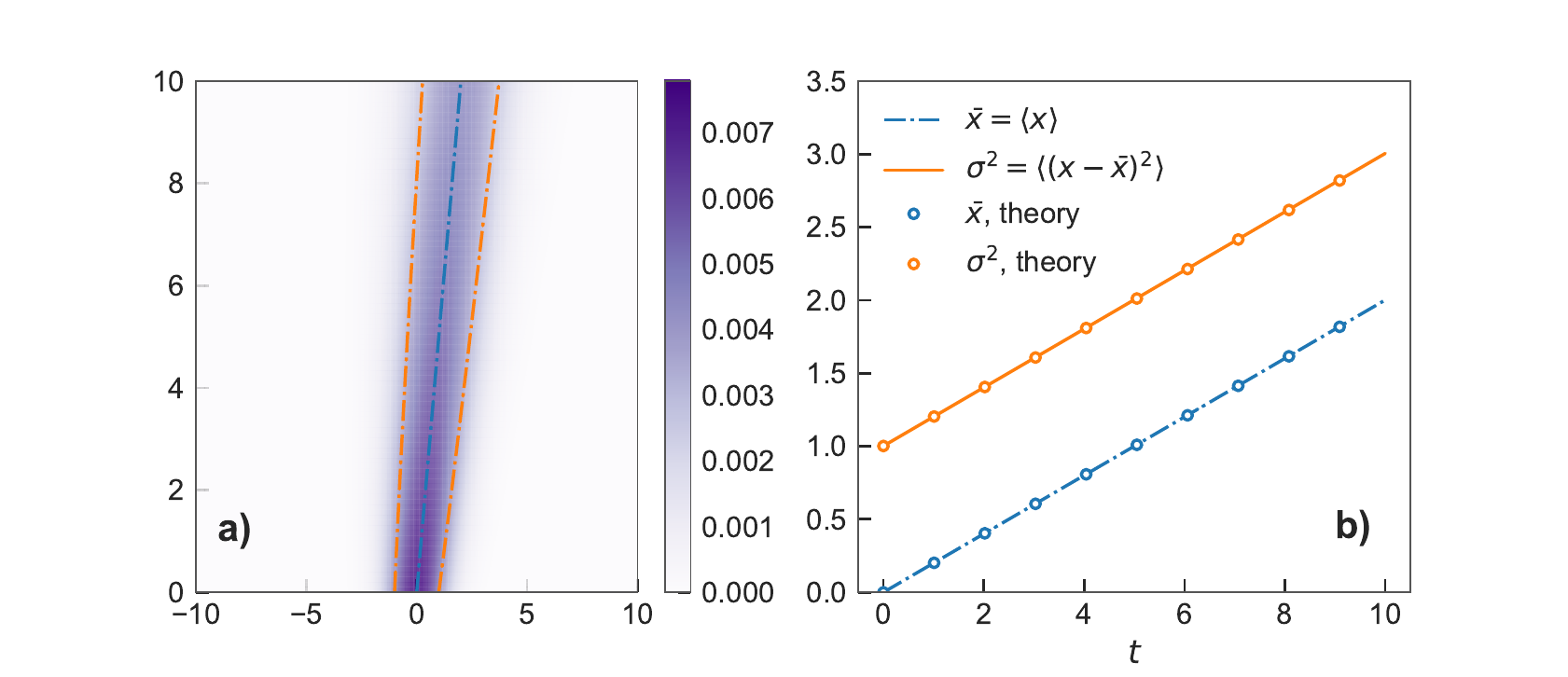}
  \caption{Solution of the Fokker-Planck equation using MPS and finite differences. (a) Density plot of the probability distribution evolved with the Fokker-Planck equation~\eqref{eq:FP} and algorithm~\eqref{eq:implicit-step}. In dashed lines we plot the expected value $\bar{x}(t)$ and the curves $\bar{x}(t)\pm\sigma(t).$ (b) Center $\bar{x}(t)$ and variance $\sigma^2(t)$ of the a probability distribution. We use $D=0.1,\mu=0.2$ and a 10-qubit discretization (1024 points), with a time step $\delta{t}=0.01.$ In circles we show the theoretical predictions. }
  \label{fig:FP}
\end{figure}

Let us assume that we have the MPS representation of the initial value $p(x,0)$ and we need to estimate the evolution of this probability distribution at later times, $p(x,t).$ Following Sect.~\ref{sec:finite-differences}, we write a finite-difference approximation to the Fokker-Planck model, replacing derivative operators with ladder operators. This transforms the Fokker-Planck equation into a first order differential equation
\begin{equation}
  \label{eq:FP-linear}
  \partial_t \ket{p(t)} = \hat{G} \ket{p(t)},
\end{equation}
generated by the linear operator $\hat{G},$ whose action we approximate with an MPO.

For concreteness, we will use the combined first-second order approximation from~\eqref{eq:finite-differences-example}
\begin{equation}
  \hat{G} = -\mu\frac{\hat{S}^+-\hat{S}^-}{2\delta{x}}+D\frac{\hat{S}^++\hat{S}^--1}{\delta{x}^2}.
\end{equation}
We also need to build an implicit integration method that works around the fact that equation~\eqref{eq:FP} is not unitary and has the potential to develop exponentially growing numerical instabilities. We have chosen a second order implicit method,
\begin{equation}
  \partial_t \ket{p(x,t)} \simeq \frac{1}{\delta t}\left[ \ket{p(x,t+\delta)} - \ket{p(x,t)}\right] \simeq \frac{1}{2}\hat{G}\left[ \ket{p(x,t+\delta)} - \ket{p(x,t)}\right].
\end{equation}
which translates into our integration recipe
\begin{equation}
  \ket{p(t+\delta t)} \simeq \left( 1 - \frac{1}{2}\delta t \hat{G} \right)^{-1}\left( 1 + \frac{1}{2}\delta t \hat{G} \right)\ket{p(x,t)}.\label{eq:implicit-step}
\end{equation}
Since $\hat{G}$ has a simple representation in terms of MPO's, we build our algorithm around the repetition of two elementary steps. First, compute the MPS for the product $\ket{\phi_1(t)}=(1+\frac{1}{2}\delta t \hat{G})\ket{p(x,t)},$ using standard simplification techniques to get the simplest and best approximation. Second, estimate the MPS $\ket{\phi_2(t)}$ that best approximates the equation
\begin{equation}
  \left( 1-\frac{1}{2}\delta{t}\hat{G} \right)\ket{\phi_2(t)} = \ket{\phi_1(t)}.
\end{equation}
In practice, we implement this step using a conjugate gradient method\footnote{One could also write quadratic optimization techniques that minimize the distance between both states, as in DMRG's \textit{correction vector techniques}~\cite{ramasesha1997,kuhner1999,schollwock2005}.}, that only requires the matrix-vector multiplication between the MPO $(1-\frac{1}{2}\delta{t}\hat{G})$ and a state represented in MPS form\ \cite{garcia-ripoll2019,garcia-ripoll2020}.

In this form, a single evolution step has a cost $\mathcal{O}(T_\text{cgs}\times \text{Simp}_{Nm})$ given by the number of iterations in the conjugate-gradient method $T_\text{cgs}$ times the cost of a single MPO-MPS multiplication, which is dominated by the simplification cost $\mathrm{Simp}_{Nm}\sim \mathcal{O}(\log_2(1/\varepsilon_\text{int})N\chi^3).$ This cost has to be compared with the classical one, in which we represent the discretized function as a vector with $2^{Nm}$ components and encode $\hat{G}$ as a sparse matrix. In this case the conjugate-gradient method will take the same number of steps, but the matrix-vector multiplication cost can be exponentially larger, roughly proportional to the size of the vector $\mathcal{O}(2^{Nm})\sim \mathcal{O}(\varepsilon_\text{int}^{-N}),$ thanks to the sparse representation. Once more, if the MPS representation remains smooth and low-entangled, with bounded $\chi,$ there can be an exponential gain.

As illustration, Fig.~\ref{fig:FP}a shows the solution of the Fokker-Planck equation for a state that is initially Gaussian with variance $\sigma(0)=1$ and center $\bar{x}(0)=0,$ living in an interval $[-10,10]$ discretized with $m=10$ bits. The finite difference methods have some intrinsic limitations. The operator $\Vert{\delta{t}\hat{G}}\Vert \propto 2^{2m}$ diverges as we add more and more qubits. Our simulation is therefore very limited both in time and in the spatial discretization. This intrinsic instability is only partially cured with the implicit methods, but not always and also not for very long simulations. If we wish to have higher precision and numerical stability, we need to develop slightly better techniques, described in the following section.

\subsubsection{Spectral split-step method}
\label{sec:time-evolution-qft}

\begin{figure}[t!]
  \centering
  \includegraphics[width=\linewidth]{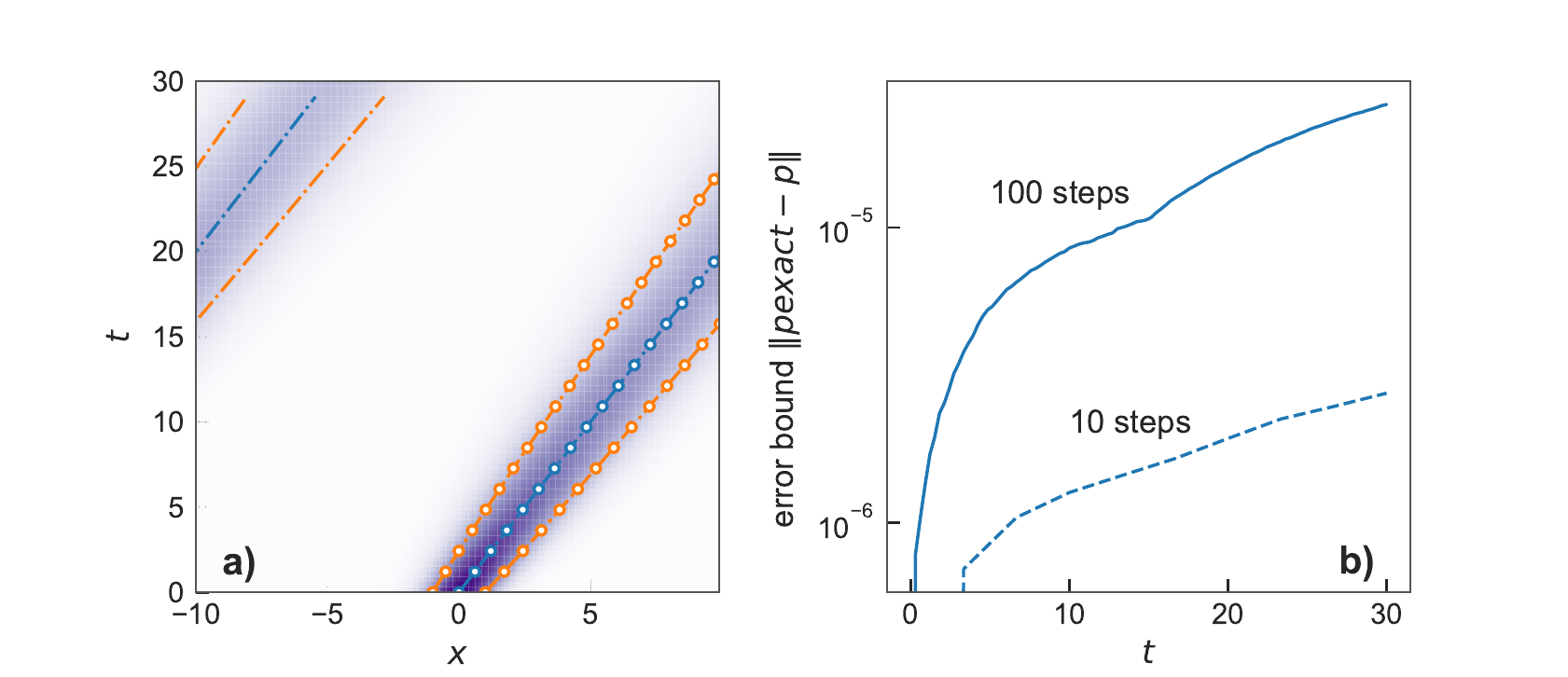}
  \caption{Solution of the Fokker-Planck equation using the Quantum Fourier Transform, for $D=0.1,\mu=0.5$ and a 14-qubit discretization (16384 points in space). (a) Density plot of the probability distribution. In dashed lines we plot the expected value $\bar{x}(t)$ and the curves $\bar{x}\pm\sigma.$ Note how the QFT implements periodic boundary conditions and the wavefunction wraps around the borders. (b) Error bounds for a time-step $\delta{t}=0.3$ (solid) and $\delta{t}=3.$ Note how the estimate decreases, indicating that our error bounds are pessimistic and that we can implement evolution for arbitrary long times without truncation errors.}
  \label{fig:FP-QFT}
\end{figure}

The spectral methods, and in particular the Fourier transform and the split-step method techniques~\cite{weideman1986}, have been traditionally used in many nonlinear Optics and quantum mechanical problems, due to their efficiency, stability and accuracy. The method is optimally designed to solve equations of the form
\begin{equation}
  \partial_t p(x,t) = G(\partial_x)p(x,t),\label{eq:general-pde}
\end{equation}
where $G(\partial_x)$ is a function of the differential operator $\partial_x$ and the coordinates are defined over a regular interval. It works by moving to Fourier space, where the generator of the evolution is a function of the momentum $k$
\begin{equation}
  \partial_t \tilde{p}(x,t) = G(ik)\tilde{p}(x,t)\; \Rightarrow \; \tilde{p}(x,t) = e^{G(ik)t}\tilde{p}(x,0).
  \label{eq:Fourier-continuous-solution}
\end{equation}
Here $\tilde{p}$ is the Fourier transform of the original function, $\tilde{p} = \hat{\mathcal{F}}p,$ over the real line.

In our discrete scenario with uniformly sampled, regular intervals, the spectral method can be implemented using the Quantum Fourier Transform~\eqref{eq:QFT} and the interpolation techniques from Sect.~\ref{sec:QFT-derivative}. The approximate solution to~\eqref{eq:general-pde} is expressed as
\begin{equation}
  \ket{p^{(m)}(t)} = \hat{\mathcal{F}}^{-1} \exp\left[ G(i\hat{k})t \right] \hat{\mathcal{F}}\ket{p^{(m)}(0)},
  \label{eq:QFT-solution}
\end{equation}
The function $G(i\hat{k})$ quasi-momentum operator~\eqref{eq:momentum} is approximated by small MPO's, as explained in Sect.~\ref{sec:mps-approximation}. Unlike the finite difference method, if the generator $G(\partial_x)$ does not depend on $x,$ the solution~\eqref{eq:QFT-solution} is exact: it works for all times in a single step.

We have implemented the recipe~\eqref{eq:QFT-solution} using MPS and MPO's. We represent $\hat{\mathcal{F}},$ $\hat{\mathcal{F}}^{-1}$ and $\exp(-\mu \hat{k} + D\hat{k}^2)$ using $3(m+1)$ MPO's of bond dimension 2. Provided that states remain weakly entangled, we can exactly solve the time evolution, without any numerical instabilities or truncation error, for any time and coefficients. As regards the efficiency, the cost is dominated by the simplification stages in the QFT and MPO-MPS product stages, which for the operators at hand seem have same practical asymptotic behavior $\mathcal{O}(Nm\times \mathrm{Simp}_{Nm})\sim \mathcal{O}(\log_2(1/\varepsilon_\text{int})^2N^2\chi^3).$ In contrast, the cost of implementing this algorithm classically, using a traditional vector representation, scales linearly with the size of the discretized function. More precisely, the FFT\ \cite{press1997} dominates the cost $\mathcal{O}(Nm 2^{Nm})\sim\mathcal{O}(\log_2(1/\varepsilon_\text{int})\varepsilon_\text{int}^{N}),$ and we expect useful speedups for many well-behaved problems.

As illustration, Fig.~\ref{fig:FP-QFT} reproduces once more the solution of the Fokker-Planck equation for a state that is initially Gaussian with variance $\sigma(0)=1$ and center $\bar{x}(0)=0,$ living in an interval $[-10\sigma,10\sigma]$ discretized with $m=14$ bits (16384 points in space). We can afford larger number of qubits and longer times than in the finite difference method, because the algorithm is orders of magnitude more efficient and very stable numerically. Notice also how the algorithm implements periodic boundaries by default; this is a feature that is very useful to avoid boundary reflections and simulate in small intervals the dynamics of propagating fields.

This recipe can be extended to problems that include dependencies on both the spatial derivatives and the spatial coordinates. We explicitly refer to equations of the form
\begin{equation}
  \partial_t p(x,t) = \left[ G(\partial_x)  + V(x)  \right] p(x,t),
\end{equation}
or higher-dimensional equivalents. The solution to this problem is no longer exact, but relies on a Trotter-Suzuki expansion of the generator
\begin{align}
  \ket{p^{(m)}(t+\delta{t})} &= e^{\left[ G(\partial_x)+V(x) \right]\delta{t}}\ket{p^{(m)}(t)}\notag\\
  &\simeq e^{V(x)\delta{t}/2}
  \hat{\mathcal{F}}^{-1}e^{G(i\hat{k})\delta{t}}\hat{\mathcal{F}}
  e^{V(x)\delta{t}/2}\ket{p^{(m)}(t)}+\mathcal{O}(\delta{t}^2),\label{eq:split-step}
\end{align}
This technique is known as a split-step method because it combines evolution steps in real space, with stages that are implemented in Fourier space\ \cite{weideman1986}. All ingredients in this formula\ \eqref{eq:split-step} are well known from earlier pages.

\subsection{Construction of the MPS}
\label{sec:exponential}

The algorithms from previous sections assume that we already have an MPS wavefunction to begin with. We will now discuss different ways in which to obtain such initial conditions.

\paragraph{Exact discretization}

The trivial way to construct an MPS representation is to begin with the discretized function $p^{(m)}(s_1,s_2\ldots s_m)$ and perform a sequential Schmidt decomposition. This has an exponential cost $\mathcal{O}(2^{3m/2})$ and fails when the number of qubits goes above 30-something, due to memory and time constraints, as we have seen in Sect.~\ref{sec:numerics-2d}.

\paragraph{Explicit formulas} Some probability distributions admit analytical expressions. We can see those functions $p(x)$ as a product $p(x)e(x)$ with the uniform distribution $e(x),$ an operation that can be approximated using the techniques from Sect.~\ref{sec:multiplication}.

An important example is the Gaussian probability distribution. When we express the coordinates in terms of qubits, the Gaussian~\eqref{eq:Gaussian-nd} becomes the partition function of an Ising model $H$
\begin{equation}
  p(x_{s_1},x_{s_2}\ldots x_{s_N}) = \frac{1}{Z(\beta)}e^{-\beta H(s_1^1,s_2^1,\ldots, s_N^m)}\ket{s_1}.\label{eq:partition-function}
\end{equation}
As discussed in Sect.~\ref{sec:multiplication}, we could write the whole MPS using the exponential of a QUBO formula. In practice, this only works well when the state $\Sigma$ is not very squeezed\footnote{When there is a lot of squeezing, or when the interval size $|b-a|$ is orders of magnitude larger than the exponential width, the computer cannot represent the extreme values that appear in the MPO tensors~\eqref{eq:QUBO-exponential-tensor}.}. More generally, we need to reconstruct the whole probability state a progressive refinement of the uniform distribution
\begin{equation}
  p(x) = \left[ \frac{1}{Z(\beta/K)} e^{-\beta H/K} \right]^K e(x),
\end{equation}
with each step implemented by an MPO
\begin{equation}
  \ket{p} = \hat{Z}_\text{MPO}(\beta/K) \cdot \hat{Z}_\text{MPO}(\beta/K)
  \cdots \hat{Z}_\text{MPO}(\beta/K) \ket{e},
  \label{eq:product}
\end{equation}
that is a better behaved Gaussian function. We have used this technique to reconstruct the MPS representations of the 2D~\eqref{eq:covariance2d} and 3D~\eqref{eq:covariance3d} Gaussian states. Figs.\ \ref{fig:entropies2d}b and \ref{fig:entropies3d}a show the maximum entanglement entropy over all bipartitions of the resulting MPS state, while Table~\ref{tab:state-sizes} discusses the amount of memory and the size of the tensors.

\paragraph{Imaginary time evolution}

Equation~\eqref{eq:product} is nothing but a discrete approximation to the imaginary time evolution of the unnormalized state $\ket{p_u}$
\begin{equation}
  \partial_\beta\ket{p_u} = - \hat{H}\ket{p_u},
\end{equation}
with uniform initial condition $\ket{p_u(\beta=0)}=\ket{e}.$ This equation can be solved for any partition function distribution~\eqref{eq:partition-function} generated by an operator $\hat{H}$ with an efficient MPO representation. This approach is reminiscent of how thermal state density matrix have been simulated using the MPS formalism~\cite{verstraete2004}, but it is much simpler, since we do not introduce any auxiliary \textit{purification} degrees of freedom.

\paragraph{Machine learning}

The challenge of writing an MPS representation of a function $p(x)$ is comparable to the challenge of constructing that function in a quantum computer: there are efficient protocols, but we do not know them a priori. The quantum computing field has developed different strategies to address this problem. A promising one is to approximate the state $\ket{p}$ using a parameterized quantum circuit that is trained using the techniques of machine learning, such as generative adversarial networks~\cite{zoufal2019}. These techniques can be extended to our domain. Instead of using a quantum-classical approach with a circuit that generates the probabilities and a neural network that discriminates, we can use the MPS as generator and apply similar training techniques.

\subsection{Porting back to the quantum computers}
\label{sec:qc-algorithms}

This work has introduced various classical MPS algorithms that use the tools of quantum computing---quantum register representation, function encodings, quantum gates and algorithms---to solve efficiently various numerical analysis tasks. These techniques, and the whole line of research, can feed back to the world of quantum computing. For instance, the Fourier interpolation algorithm from Sect.~\ref{sec:Fourier-interpolation} has an immediate translation to a real quantum computer.

Other algorithms require some fine tuning. In linear interpolation and finite differences, the the $\hat{S}^\pm$ operator was designed to use irreversible arithmetic. However, we know that a similar operator can be implemented or approximated using auxiliary qubits and reversible arithmetic---half and full adders---, or other ideas from quantum simulation\ \cite{cao2013}.

Something similar happens in the case of PDE solvers. Our discussion has focused on the non-unitary evolution induced by Fokker-Planck techniques, studying the implementation of the non-unitary operator $\exp[G(\hat{k})].$ Algorithms\ \eqref{eq:QFT-solution} and \eqref{eq:split-step} can be trivially generalized to solve Schrödinger equations such as
\begin{equation}
  \partial_t \psi(x,t) = \left[-i\frac{\hbar}{2m}\nabla^2  -i V(x)/\hbar \right]\psi(x,t).
\end{equation}
The generator of this equation is anti-Hermitian and both $\exp[\delta{t}\,G(i\hat{k})]$ and $\exp[-iV(x)\delta{t}/\hbar]$ can be implemented as a unitary gate in the quantum register. As in the MPS case, the scaling of the algorithm is problem dependent. The exponential $\exp[i\alpha\hat{k}^2]$ can be implemented with $\mathcal{O}(N^2m^2)$ steps, but the exponential of $V(x)$ may have a more complicated scaling, strongly dependent on function arithmetic. However, we expect that the smoothness of usual potentials will also lead to simple approximations with quasi-local gates.

\subsection{Implementation}
\label{sec:implementation}

Together with this manuscript, we have developed an accompanying MPS library\ \cite{garcia-ripoll2019} and supplementary driving code\ \cite{garcia-ripoll2020}, all of which are available as separate Zenodo/GitHub repositories. The SeeMPS library\ \cite{garcia-ripoll2019} contains an implementation of all MPS fundamental algorithms, including the fundamental representation, approximation of linear combinations of states, MPO-MPS contractions, conjugate-gradient for matrix inversion, and Quantum Fourier Transform. The library is written using the literate-programming approach, and combines Python code with the mathematical formulation of the algorithms that we have explained.

The supplementary material to this manuscript\ \cite{garcia-ripoll2020} is also available as Zenodo/GitHub respository. It builds on the given version of the SeeMPS library, using it to implement all methods and simulations in this work . In particular, it shows (i) how to construct the quantum-register approximations of the probability distributions studied in Sect.\ \ref{sec:numerics-1d}, (ii) efficient techniques to construct Gaussian multivariate probability distributions using MPS, (iii) a study of Fourier interpolation with MPS, (iv) a program to solve the Fokker-Planck using the finite-difference and conjugate-gradient methods, and (v) a solution of the Fokker-Planck equation using the MPS split-step method. Those codes are the ones that have been used to produce the plots in this work.

\section{Discussion and outlook}
\label{sec:outlook}

This work has presented many numerical algorithms for constructing, manipulating and interrogating multivariate functions in MPS-simulated quantum registers. We have shown that, heuristically, the renormalization provided by the quantum register representation is key to the creation of states with low entanglement, capable of encoding smooth, differentiable functions.

The use of tensor network states is a modern development in numerical analysis~\cite{grasedyck2013,bachmayr2016}. In the field of \textit{low-rank tensor approximations}, a multivariate function $\psi(x_1,x_2\ldots x_N)$ is approximated by a contraction of tensors, labeled either by the continuous variables $x_i,$ or discretized versions of them $x_{s_i}.$ In other works, the spatial degrees of freedom are replaced by labels in some local mode expansion~\cite{iblisdir2007}. These approaches use a Hilbert space decomposition that preserves the notion of local degrees of freedom, by expanding a function in a normal basis of modes
\begin{equation}
  f(x_1,x_2,\ldots,x_N) = \sum c(i_1,i_2,\ldots,i_N) \phi_{i_1}(x_1)\phi_{i_2}(x_2)\cdots\phi_{i_N}(x_N),
\end{equation}
and using a low-rank tensor approximation to represent $c(i_1,i_2,\ldots,i_N).$ It can be argued that such encodings are isomorphic to the A-order tensor structure discussed in Section\ \ref{sec:numerics-2d}. Since the orthonormal modes $\phi_{i_n}(x_n)$ can be expanded in the basis of $m$ qubits that represents the $n$-th coordinate, $\ket{s_{n}^1,\ldots,s_{n}^m},$ these modes can be regarded as an isometry from the space of qubits $\{s_n^1,\ldots,s_n^m\}$ to the mode coordinate $\{i_n\}.$ This isometry cannot cure the fact that the entanglement between coordinates grows in an unfavorable way with the dimensionality of the problem, as we saw in Section\ \ref{sec:numerics-2d}. This entanglement growth will affect the complexity of any splitting of the tensor $c(i_1,i_2,\ldots,i_N)$ just as badly as it affected our MPS representation.

In this work we have presented a quantum register discretization that solves those limitations. In this representation each local coordinate is exploded into $m$ qubits of information,  each of them probing the function at a different length scale. As we have seen in Sect.~\ref{sec:numerics-2d}, these qubits need not be kept together, and there may be more efficient renormalization schemes when we group common length scales that are more strongly correlated. Returning to the mode representation mentioned before, this technique of grouping qubits is comparable to a multi-scale wavelet modal representations
\begin{equation}
  f(x_1,x_2,\ldots,x_N) = \sum c(i_1,i_2,\ldots,i_N) \prod_n \phi_{i_n}(x_1,x_2,\ldots,x_N),
\end{equation}
where where each $\phi_{i_n}(x_1,x_2,\ldots,x_N)$ represents the oscillations of the function at a different length scale, and our tensor $c(i_1,\ldots,i_N)$ explores correlations between different length scales. This implicit renormalization scheme is a discovery of the quantum computing community that has not been sufficiently exploited so far and which may empower many recent developments in the field of quantum finance and probability distribution analysis~\cite{rebentrost2018, orus2019, woerner2019}.

The algorithms in this paper bend themselves to a broad family of problems which have been considered in the quantum computing world. We have illustrated the solution of time-dependent partial differential equations, but the same techniques can be extended to stationary problems. This way, the MPS-simulated quantum register becomes a natural tool to solve the Poisson equation~\cite{cao2013}, the wave equation~\cite{arrazola2018}, the fluid equations~\cite{steijl2018}, or even the Schrödinger equation itself. This implies not just abstract, fundamental studies in Physics, but practical applications in fields such as aerodynamics or finance. We expect new applications of quantum-inspired finance that reach beyond the state-of-the-art~\cite{orus2019}, providing new schemes for evaluating financial products~\cite{rebentrost2018}, performing risk analysis~\cite{woerner2019} and even more sophisticated time-dependent simulations and tracking.

Let us also remark that the algorithms developed in this work are of a heuristic nature. All methods and techniques in Sections~\ref{sec:quantum-inspired-algorithms} and \ref{sec:analysis} assume states with low entanglement. This approximation is bound to break at some point, either because of functions with broad spectra and complex structure, or because of increased dimensionality. Quantum computers become a valuable tool that still has an exponential advantage over classical algorithms, and which may profit from the ideas and developments associated to MPS-simulated quantum register techniques, as discussed before [cf. Sect.~\ref{sec:qc-algorithms}].

Finally, all simulations in this paper have been implemented with the SeeMPS Python library~\cite{garcia-ripoll2019,garcia-ripoll2020}. While Python is a slow language, the algorithms are high-level and benefit from the efficient tensor operations that are available in Numpy and Scipy. This makes the code performant enough for practical applications. However, it is undeniable that recent frameworks based on TensorFlow~\cite{mislted2019} and highly parallelized architectures for tensor contractions~\cite{huang2019} would be a better platform to real-world deployment.

J.J.G.-R. thanks Andrea Cadarso for discussions around the motivation of the method and its applications. This material is based upon work supported by the Air Force Office of Scientific Research under award number FA2386-18-1-4019. This work has been supported by funding from project PGC2018-094792-B-I00 (MCIU/AEI/FEDER, UE) and and CAM/FEDER Project No. S2018/TCS-4342 (QUITEMAD-CM).


\end{document}